\documentclass{article}
\usepackage[toc,page]{appendix}
\usepackage{authblk}
\usepackage{orcidlink}
\usepackage{graphicx}
\usepackage{booktabs}
\usepackage{makecell}
\usepackage{microtype}
\usepackage[backend=biber,natbib=true,
autocite=inline, citestyle=numeric-comp,
isbn=false,block=space,date=short,
maxbibnames = 10, bibencoding=utf8, sorting = none]{biblatex}
\bibliography{bib.bib}
\AtEveryBibitem{\clearfield{month}}
\AtEveryBibitem{\clearfield{pages}}
\DeclareFieldFormat{title}{#1}
\DeclareFieldFormat{booktitle}{#1}
\DeclareFieldFormat{citetitle}{#1}
\DeclareFieldFormat{journaltitle}{#1}
\usepackage[left=18mm,right=18mm,bottom=18mm,top=18mm]{geometry}
\usepackage{amsmath}
\usepackage{amssymb}
\usepackage{braket}
\usepackage{bbm}
\usepackage{mathtools}
\usepackage{xcolor}
\usepackage{csquotes}
\usepackage{caption}
\usepackage{subcaption}
\usepackage{hyperref}
\hypersetup{
    colorlinks=true,
    linkcolor=blue,
    filecolor=magenta,      
    urlcolor=cyan,
    pdftitle={Overleaf Example},
    pdfpagemode=FullScreen,
    }
\usepackage{parskip}
\usepackage{svg}
\usepackage[most]{tcolorbox}
\usepackage{nicefrac}
\usepackage[normalem]{ulem}
\usepackage{quantikz}

\definecolor{steelblue}{rgb}{0.27, 0.51, 0.71}
\definecolor{firebrick}{rgb}{0.7, 0.13, 0.13}
\definecolor{greenish}{rgb}{0, 0.7, 0}

\tcbset{highlight math style={enhanced,
  colframe=orange!90!white,colback=orange!15!white,boxrule=1pt}}
\newcommand {\tr} {\mathrm{tr}}
\newcommand {\ketbra}[2] {\lvert #1 \rangle \langle #2 \rvert} 
 
\newcommand {\norm} [1] {\lvert \lvert #1 \rvert \rvert} 
\newcommand {\expectation} [1] {\langle  #1 \rangle}

\title{Computing Statistical Properties of Velocity Fields\\ on Current Quantum Hardware}

\author[1,2]{Miriam Goldack \orcidlink{0009-0009-3624-721X}}
\author[3]{Yosi Atia}
\author[3]{Ori Alberton}
\author[1,4]{Karl Jansen}

\affil[1]{Deutsches Elektronen-Synchrotron DESY, Platanenallee 6, 15738 Zeuthen, Germany}
\affil[2]{Institut für Strömungsmechanik und Technische Akustik (ISTA), Technische Universität Berlin, Straße des 17. Juni 135, 10623 Berlin, Germany}
\affil[3]{Qedma Quantum Computing, Rokach Blvd 101, Tel Aviv, Israel}
\affil[4]{Computation-Based Science and Technology Research Center, The Cyprus Institute, 20 Kavafi Street, 2121 Nicosia, Cyprus}

\date{December 29, 2025}

\begin{document}

\maketitle

\section*{Abstract}
Quantum algorithms are gaining attention in Computational Fluid Dynamics (CFD) for their favorable scaling, as encoding physical fields into quantum probability amplitudes enables representation of $2^n$
 spatial points with only $n$ qubits. A key challenge in Quantum CFD is the efficient readout of simulation results, a topic that has received limited attention in literature. This work presents methods to extract statistical properties of spatial velocity fields, such as central moments and structure functions, directly from parameterized ansatz circuits, avoiding full quantum state tomography. As a proof of concept, we implement our approach for 1D velocity fields, encoding 16 spatial points with 4 qubits, and analyze both a sine wave signal and four snapshots from Burgers' equation evolution. Using Qedma's error mitigation software QESEM, we demonstrate that such computations achieve high accuracy on current quantum devices, specifically IBMQ's Heron2 system \texttt{ibm\_fez}.

\textbf{Keywords:} Quantum Computational Fluid Dynamics (QCFD), Quantum Error Suppression and Error Mitigation (QESEM), Readout Problem, Turbulence, Statistical Properties, Burgers' Equation

\section{Introduction}
Fluid dynamical systems are governed by partial differential equations (PDEs), such as the Navier-Stokes equation or its simplified form, the Burgers' equation \cite{BURGERS1948171}. Due to their inherent nonlinearity and sensitivity to initial conditions, analytical solutions are often unavailable, making numerical simulations essential. However, achieving high accuracy simulations requires resolving fine-scale structures through a fine spatial discretization with a large number of grid points, especially in turbulent flows, intricate geometries, or three-dimensional systems. This significantly increases memory and processing requirements, posing a major challenge for classical Computational Fluid Dynamics (CFD). 

The high computational cost of classical fluid simulations has led to a growing interest in quantum computing  fluid dynamics (QCFD) algorithms. Linear PDEs can be addressed using Quantum Linear System Algorithms as summarized in \cite{morales2025}. Starting with the Harrow-Hassidim-Lloyd algorithm \cite{HHL} in 2009, which was the first quantum algorithm for solving systems of linear equations with a runtime that scales logarithmically in the input matrix dimension, there have been several advancements in that domain \cite{Berry_2017, Childs_2021, Bravo_Prieto_2023, jennings_2023, jennings2024cost, Chen_2024}. 
The Coherent Variational Quantum Linear Solver has even been used to solve the nonlinear Burgers' equation in \cite{Bosco_2024}. Other approaches to address nonlinear PDEs include transformations to ordinary differential equations as shown in \cite{Gaitan2020, Oz2021, Liu_2021}, Lattice Boltzmann Methods \cite{li_2024}, or the Cole-Hopf transformation for the Burgers' equation \cite{uchida2024}. Variational Quantum Algorithms (VQA), as presented in \cite{Lubasch_2020, jaksch2022, Siegl_2025}, are based on classical discretization and time integration methods and treat the solution of the equation as a variational problem, where the cost function is evaluated on the quantum computer. This allows to solve various PDEs by adjusting the cost function to the given equation, boundary conditions and external forcing \cite{Over_2025}.   

Most of these algorithms represent entire fields within quantum states using amplitude encoding: in an $n$-qubit system, each of the $2^n$ basis states encodes a spatial location, with its amplitude representing a physical quantity like velocity. This compact encoding could, in principle, enable vastly higher spatial resolutions than classical methods, and offers a promising path towards simulating complex fluid systems beyond classical limits \cite{jaksch2022}. 

While quantum algorithms offer theoretical speedups for solving fluid dynamics problems, two major challenges remain: preparing and verifying suitable quantum states, and extracting meaningful results from them.
State preparation can be straightforward in simple cases, such as initializing a delta function or a Gaussian distribution where efficient methods are known \cite{grover2002, kitaev2009}. For more complex initial conditions, recent approaches construct quantum circuits by leveraging tensor network representations of the target state \cite{Ran2020, Malz2024, Iaconis2024, Smith2024}. In some cases, this has enabled circuit depths that are several orders of magnitude shorter than those required by general exponential loading schemes, suggesting that many classical datasets may be more efficiently loadable than previously assumed \cite{Jumade_2023}. Despite these improvements in state preparation, the main conceptual obstacle for practical quantum simulation algorithms of fluid dynamics remains the extraction of relevant physical quantities from the quantum state.

As Aaronson \cite{Aaronson2015} pointed out, quantum algorithms typically output a quantum state rather than the full solution vector. Retrieving detailed information from this state, such as individual velocity values, generally requires repeated measurements, scaling exponentially with system size, thereby undermining not only the potential computational speedup but also the exponential memory efficiency that makes such encodings attractive in the first place. Full state tomography becomes infeasible for large systems, as reconstructing an $n$-qubit density matrix requires estimating $4^n-1$ independent real parameters via repeated measurements in different bases. Although alternative readout methods (such as compressed sensing~\cite{Cramer2010, Flammia_2012}, MPS tomography~\cite{Lanyon_2017}, Neural Quantum States~\cite{lange2024}, or shadow tomography~\cite{aaronson2018} and classical shadows~\cite{Huang_2020}) can significantly reduce the measurement cost compared to full tomography, they are typically only approximate or rely on specific assumptions. These assumptions include, for example, low-rank structures or restricted entanglement, which may not hold in general CFD applications and can render such methods infeasible at larger scales. A similar limitation applies to tensor-network methods, which are often considered direct competitors to quantum approaches: their output is likewise a full quantum state, meaning that they ultimately face the same tomography bottleneck when full state information is requested.

In many practical scenarios, especially in turbulence research, full information about the velocity field is not required. Instead, key statistical properties are often of primary interest \cite{Frisch_1995, Pope_2000}. If these could be computed directly on the quantum computer, the need for exponential-cost state tomography would be avoided, and post-processing could be significantly simplified. Supporting this idea, recent work by Uchida et al.~\cite{uchida2024} proposes a quantum algorithm for solving the Burgers' equation that enables the efficient extraction of multi-point functions from the quantum state without full tomography, assuming access to fault-tolerant quantum hardware.

This study pursues a similar idea to Uchida et al.~\cite{uchida2024}, in the sense that here too, statistical properties of the quantum state are extracted without full state tomography. In contrast to their approach, we develop a method that is applicable to currently available hardware, and demonstrate it on IBM quantum processing units (QPUs). Instead of reconstructing the full quantum state via tomography, we directly compute key spatial statistics of velocity fields that are encoded in the amplitudes of a quantum state vector. To this end, we assume that a quantum algorithm solves the underlying PDE and encodes the spatial distribution of the velocity at each time step in the quantum state using amplitude encoding. In our implementation, this state is prepared using a VQA, following a parameterized ansatz circuit similar to~\cite{Lubasch_2020, jaksch2022, Siegl_2025}. Although we demonstrate our method on the example of Burgers' turbulence, the concept is applicable to other equations or other contexts in general.

All computations are performed on IBMQ's Heron2 system \texttt{ibm\_fez}, a 156-qubit quantum computer, using Qedma's error suppression and error mitigation software QESEM \cite{QESEM} (see Appendix \ref{sec:QESEM}). Our results show that even when running on noisy quantum hardware, the extraction of these few, yet informative, statistical values is feasible using error mitigation. This significantly reduces the demands on readout and post-processing, and provides a practical pathway toward meaningful quantum simulations on near-term quantum devices. 

Section~\ref{sec:background} introduces Burger's equation, turbulence statistics and amplitude encoding. Building on this, we develop a tailored measurement strategy for extracting the components for central moments and structure functions from quantum states in Section~\ref{sec:measurement_strategy}. This is followed by Section~\ref{sec:circuit_design} on circuit design, where we present hardware-aware modifications to optimize execution on IBM’s heavy-hex devices. The subsequent Section~\ref{sec:hardware_results} presents results from measurements on real quantum hardware. Finally, we summarize our findings and provide an outlook on future developments in Section~\ref{sec:discussion}.

\section{Background}\label{sec:background}
\subsection{Turbulence Statistics}\label{sec:turbulence_statistics}

The one-dimensional viscous Burgers' equation \cite{BURGERS1948171} is a nonlinear partial differential equation commonly used as a simplified model to study turbulence and nonlinear wave propagation. It reads:
\begin{equation}
    \frac{\partial u}{\partial t} + u \frac{\partial u}{\partial x} = \nu \frac{\partial^2 u}{\partial x^2} + f(x,t)
\end{equation}
where $u(x,t)$ is the velocity field, $\nu$ is the kinematic viscosity, and $f(x,t)$ represents an external forcing term. Despite its simplicity, the equation captures essential features of more complex fluid dynamics systems, such as the formation of shock waves, steep velocity gradients, and energy cascades that are typical in turbulent flows \cite{Frisch_1995,Pope_2000}. Previous studies have demonstrated that rare events and anomalous scaling behavior in the Burgers' equation can be effectively captured using statistical approaches based on path-integral and Monte Carlo formulations \cite{Mesterhazy_2011, Margazoglou_2019}. This motivates our focus on extracting statistical properties, that characterize flow structures and turbulence, rather than reconstructing the full velocity field.

To this end, we discretize the spatial domain into $N$ points and define the spatial mean as
\begin{equation}
    \langle u \rangle = \frac{1}{N}\sum_{i=1}^N u_i.
\end{equation}
Using this, one important class of statistical descriptors, the central moments of the spatial velocity distribution at a fixed time step, can be defined as
\begin{equation}
\begin{split}
    \langle u'^k \rangle &= \frac{1}{N} \sum_{i=1}^N (u_i - \langle u \rangle)^k .\label{eq:central_moments_definition} 
\end{split}
\end{equation}
These spatially averaged moments provide a compact description of spatial heterogeneity and are especially useful for identifying flow structures such as coherent vortices or regions of enhanced shear \cite{Pope_2000}.

The mean value is the average of the velocity values and the central moments refer to deviations from it. The first central moment (mean deviation) is zero by definition. The second moment (variance) quantifies how strongly the velocity field fluctuates around its mean value and is related to turbulence intensity and the kinetic energy of the fluctuating part of the flow field. Higher-order moments, particularly the third (related to the skewness) and fourth (related to the flatness or kurtosis), offer deeper insight into the asymmetry and extreme behavior of the velocity distribution. Specifically, the skewness captures the degree of asymmetry, which is often connected to anisotropies in the flow. It provides insight into the prevalence of coherent structures such as vortices or localized high-energy regions. The flatness serves as a measure of intermittency and highlights the prevalence of extreme events \cite{Frisch_1995}.

\begin{figure}[ht!]
    \centering
        \includegraphics[width=0.8\linewidth]{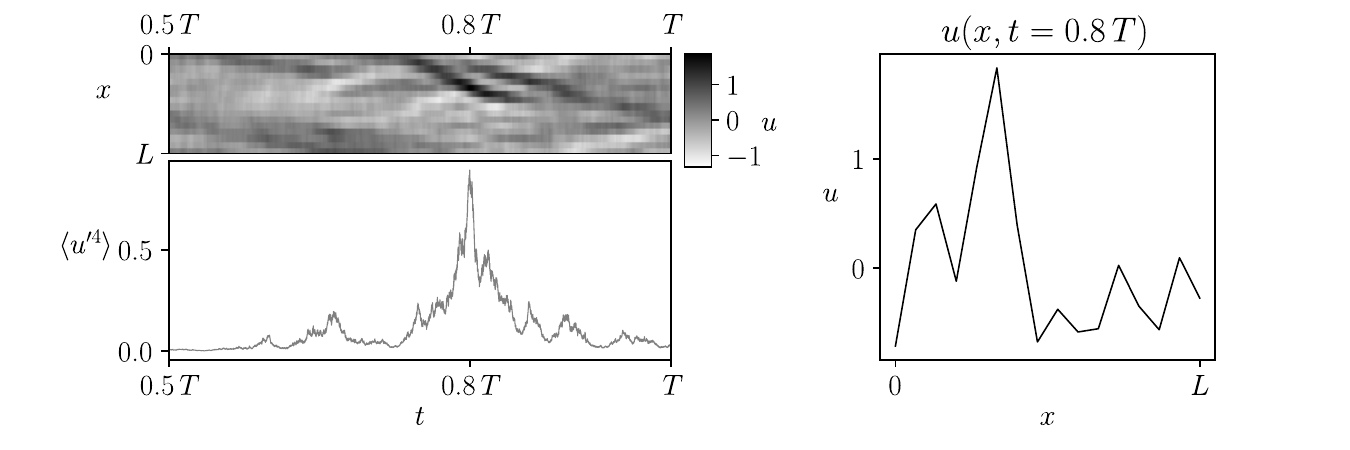}
    \caption{Top left: Example of the spatiotemporal evolution of the forced Burgers’ equation (random initial condition, $N=16$, $\nu=0.1$), showing the formation and propagation of shock waves. This panel is a subset of the full simulation shown in the top panel of Figure~\ref{fig:moments_over_time}. Bottom left: Temporal evolution of the corresponding fourth central moment, peaking at $0.8\,T$, where $T$ denotes the final simulation time. Right: Spatial velocity profile at $0.8\,T$, showing a steep velocity gradient.}\label{fig:sneak_peak}
\end{figure}

Such moments are particularly sensitive to the presence of shock-like structures, which appear in the Burgers' equation as steep gradients driven by the nonlinear advection term. These structures cause large local velocity changes, leading to enhanced skewness and flatness values. In this way, shock waves leave a statistical signature even when the full field is not explicitly reconstructed, as illustrated in Figure~\ref{fig:sneak_peak}: for a random initial condition with $N=16$ grid points, viscosity $\nu=0.1$, and stochastic forcing, a pronounced structure emerges at $t=0.8\,T$, where $T$ denotes the total simulation time (see Subsection~\ref{sec:burgers_turbulence} for full simulation details). At this point, a sharp peak in the fourth central moment coincides with a steep velocity jump in the spatial profile.

A complementary tool for analyzing characteristics such as the multi-scale nature of turbulent flows are structure functions. The $k$th-order structure function is defined as: 
\begin{equation}
    S_k(r) = \langle \vert \Delta u\vert^k\rangle_{x,t} = \langle \vert u(x+r, t) - u(x, t) \vert^k \rangle_{x,t}
\end{equation}
where $\Delta u$ is the velocity increment over a spatial separation $r$, and the angle brackets denote an average over space and time. Structure functions describe how velocity differences scale with separation distance and are particularly useful for characterizing the distribution of energy and intermittency across different length scales~\cite{Frisch_1995, Pope_2000}. They reveal correlation lengths, coherent structures and scaling properties in the velocity field, providing insight into transitions between different turbulent regimes.

By computing central moments and structure functions directly on the quantum device, we aim to characterize flow behavior in a compact, meaningful way without the overhead of full state tomography.

\subsection{Quantities of Interest}\label{sec:target_quantities}

The second, third, and fourth central moments, as defined in Equation~(\ref{eq:central_moments_definition}), can be rewritten as linear combinations of the mean $\langle u \rangle$ and the so-called raw moments $\langle u^k \rangle = \frac{1}{N} \sum_{i=1}^N u_i^k$:
\begin{align}
    \makebox[3.8cm][l]{\text{mean}}      & \langle u \rangle = \frac{1}{N} \sum_{i=1}^N u_i \label{eq:mean_lin_comb}\\
    \makebox[3.8cm][l]{\text{2nd central moment}} & \langle u'^2\rangle = \frac{1}{N} \sum_{i=1}^N u_i^2 - \langle u\rangle^2 \label{eq:2nd_cm_lin_comb}\\
    \makebox[3.8cm][l]{\text{3rd central moment}} & \langle u'^3\rangle = \frac{1}{N} \sum_{i=1}^N u_i^3 - 3 \langle u\rangle \frac{1}{N} \sum_{i=1}^N u_i^2 + 2 \langle u\rangle^3 \label{eq:3rd_cm_lin_comb}\\
    \makebox[3.8cm][l]{\text{4th central moment}} & \langle u'^4\rangle = \frac{1}{N} \sum_{i=1}^N u_i^4 - 4 \langle u\rangle \frac{1}{N} \sum_{i=1}^N u_i^3 + 6 \langle u\rangle^2 \frac{1}{N} \sum_{i=1}^N u_i^2 - 3 \langle u\rangle^4 \label{eq:4th_cm_lin_comb} 
\end{align} 

Regarding the structure functions, it is possible to compute the spatial average for a single time step as 
\begin{equation}
    \langle \vert \Delta u\vert^k\rangle_{x} = \langle \vert u(x+r) - u(x) \vert^k \rangle_{x}.
\end{equation}
To obtain the full spatiotemporal average, this value needs to be computed for each time step and then averaged.

Since the velocity values are real, these expressions for even exponents $k$ can be decomposed into linear combinations of basic algebraic sums over the data. This simplifies the calculation of structure functions, as the required terms involve only powers and products of velocity values. Thus, the computation of structure functions becomes a direct extension of the central moment calculations, with the relevant sums adjusted to the specific exponent. As an example, we show the decomposition into measurable sums for the exponents $k=2$ and $k=4$:

\begin{align}
&\langle \vert \Delta u\vert^2\rangle_{x} = \frac{1}{N} \Bigg[ \sum_{i=1}^N u_i^2 - 2 \sum_{i=1}^N u_{i+r} u_i + \underbrace{\sum_{i=1}^N u^2_{i+r}}_{=\sum_{i}u_i^2} \Bigg] = \frac{1}{N} \Bigg[ 2 \sum_{i=1}^N u_i^2 - 2 \sum_{i=1}^N u_{i+r} u_i \Bigg] \label{eq:2nd_sf_lin_comb}\\
&\langle \vert \Delta u\vert^4\rangle_{x} = \frac{1}{N} \Bigg[ 2 \sum_{i=1}^N u_i^4 - 4 \sum_{i=1}^N u_{i+r}^3 u_i - 4 \sum_{i=1}^N u_{i+r} u_i^3 + 6 \sum_{i=1}^N u_{i+r}^2 u_i^2 \Bigg]. \label{eq:4th_sf_lin_comb}
\end{align}

Estimating both the raw moments and the terms appearing in Equations~(\ref{eq:2nd_sf_lin_comb}) and (\ref{eq:4th_sf_lin_comb}) enables the reconstruction of the central moments and the associated structure functions, which constitute the primary observables of this work.

\subsection{Quantum Encoding}\label{sec:quantum_encoding}
To enable a highly compact representation of the velocity field with exponential storage efficiency compared to classical methods, we  employ amplitude encoding. In this approach, the normalized values of the velocity field are directly mapped to the probability amplitudes of quantum states, i.e.:
\begin{equation}
    \ket{\phi_u} = \frac{1}{\|u\|} \sum_j u_j\ket{j} 
\end{equation}
where $u_j$ are the elements of the velocity field after normalization, and $||u||$ is the 2-norm of $u$. This means that each basis state corresponds to a specific spatial point, with its amplitude encoding the value of the velocity field at that position. The full quantum state thus represents the spatial distribution of the velocity at a fixed point in time.

To prepare the quantum state $\ket{\phi_u}$, we use an ansatz circuit, which is a parameterized quantum circuit designed to approximate a target quantum state. The parameters of this circuit are optimized such that the resulting state closely matches the desired encoded velocity field.
The state $\ket{\phi_u}$ can be prepared by adjusting the parameters $\theta$ of the ansatz $U(\theta)$:
\begin{equation}
    \ket{\phi_u} = U(\theta) \ket{0}^{\otimes n}\label{eq:parameterized_ansatz}
\end{equation}
where $n$ is the number of ansatz qubits. Since each of the $N=2^n$ basis states of the $n$-qubit register encode one spatial position, this representation achieves an exponential storage efficiency with respect to the number of qubits.

Building on the idea of using a parameterized circuit to approximate a quantum state, QCFD methods such as the VQA~\cite{Lubasch_2020, jaksch2022, Siegl_2025} treat the time evolution of PDEs as a sequence of variational optimization problems. At each time step, the parameters of the ansatz $U(\theta)$ are re-optimized such that the residual of the underlying PDE is minimized, thus updating the state in analogy to classical time-integration schemes. 
Since quantum states are normalized by construction, while physical velocity fields are not, the overall amplitude must be recovered separately. For this purpose, the norm $\|u\|$ of the velocity field is factored out and treated as an additional scalar parameter, optimized together with the ansatz parameters. This separation ensures that the ansatz encodes the shape of the velocity distribution, while $\|u\|$ restores its correct physical magnitude. More detailed discussions of this variational formulation can be found in \cite{Lubasch_2020, jaksch2022}. 

Here, we assume that such a time integration scheme is used and the parameter optimization is done for every time step. This means that the parameters that effectively encode the system's evolution are known and can be used to reconstruct individual states. The remainder of the paper focuses on extracting the statistical properties of interest from the resulting states.

\section{Measurement Strategy}\label{sec:measurement_strategy}

The expressions for the central moments and structure functions derived in 
Equations (\ref{eq:mean_lin_comb})-(\ref{eq:4th_sf_lin_comb}) involve, respectively, sums over powers of the velocity components and higher order correlation functions. Because the quantum state is normalized, the sum of squares is trivially equal to one and does not need to be evaluated explicitly. The algorithms used to extract these statistical quantities are designed with the noise characteristics of current quantum hardware in mind, which renders advanced techniques such as amplitude estimation and block encoding ~\cite{brassard2000quantum, gilyen2019quantum} impractical. 
Instead, the statistical properties of interest are formulated as expectation values of suitable observables, enabling the use of error-mitigation strategies (specifically QESEM) to reduce noise-induced biases in the measured data.

\subsection{Estimating \texorpdfstring{$\expectation{u}$}{<u>}}

\begin{figure}[htbp]
  \centering
    \begin{center}
        \begin{quantikz}
            \lstick{$\ket{0}$} & \gate{H} & \ctrl{1} & \meter{X} \\
            \lstick{$\ket{\phi}$} & \qwbundle{} & \gate{U} & \qwbundle{} 
        \end{quantikz}
        \hspace{2cm}
        \begin{quantikz}
            \lstick{$\ket{0}$} & \gate{H} & \octrl{1} & \ctrl{1} & \meter{X} \\
            \lstick{$\ket{0}$} &\qwbundle{} & \gate{U_1} & \gate{U_2} & \qwbundle{} 
        \end{quantikz}
    \end{center}
    \captionof{figure}{\label{fig:Htest} Left: Hadamard test. The ancilla is measured in the \(X\) basis (denoted by the meter symbol), yielding the Pauli-$X$ expectation value $\textrm{Re}\langle \phi | U | \phi \rangle$. Right: Modified Hadamard test, estimating $\textrm{Re}\langle \phi_1 | \phi_2 \rangle$ where $\ket{\phi_1}, \ket{\phi_2}$ are prepared by applying $U_1,U_2$ on the all-zero state respectively.}
\end{figure}
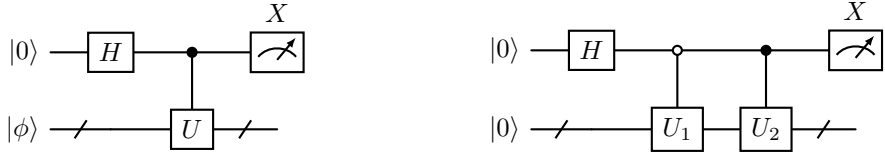

The spatial mean value \( \langle u \rangle \) of the velocity field is computed using the Hadamard test~\cite{Cleve_1998}, a versatile quantum subroutine that enables the efficient estimation of inner products and expectation values.
In its standard form, shown in Figure~\ref{fig:Htest} on the left, the Hadamard test estimates the real part of the expectation value \( \langle \phi | U | \phi \rangle \), by measuring the Pauli-$X$ observable of the top ancilla qubit: 
\begin{equation}
\langle X_a \rangle = P(+) - P(-) =  \textrm{Re} \langle \phi | U | \phi \rangle.
\end{equation}
Here, \(P(+)\) and \(P(-)\) denote the probabilities for measuring the basis states in the \(X\) basis.

This principle can be adapted to estimate the inner product between two distinct quantum states \( |\phi_1\rangle = U_1 |0\rangle \) and \( |\phi_2\rangle = U_2 |0\rangle \), using a modified Hadamard test as described in~\cite{Xiong_2024}. As shown in Figure~\ref{fig:Htest} on the right, the circuit applies \( U_1 \) under a negative control and \( U_2 \) under a positive control. In~\cite{Xiong_2024}, the real part of the overlap \( \langle \phi_1 | \phi_2 \rangle \) is obtained by measuring the ancilla in the Pauli-\(Z\) basis after a final Hadamard gate. Conceptually, this is equivalent to a direct measurement in the \(X\) basis. This yields the same quantity:
\begin{equation}
\langle X_a \rangle = \text{Re} \langle \phi_1 | \phi_2 \rangle.
\end{equation}

In our application, the quantum state $\ket{\phi_u} = U(\theta) \ket{0}$ encodes the real-valued velocity field in its amplitudes. To compute the spatial mean $\langle u\rangle$, we evaluate the overlap $\langle \phi_u \vert + \dots +\rangle$ between the velocity-encoding state and a uniform superposition state. The latter represents an equal-weighted sum over all spatial positions, $\ket{+\dots +}=\frac{1}{\sqrt{N}}\sum_i\ket{i}$. In explicit vector form, this corresponds to the inner product between the normalized velocity vector and a column vector of ones: 

\begin{equation}
    \langle u\rangle = \frac{1}{N} \sum_{i=1}^N u_i = \frac{1}{N}
    \underbrace{
    \begin{bmatrix}
        u_1 & u_2 & \hdots & u_N
    \end{bmatrix}}_{\|u\| \bra{\phi_u}} 
    \underbrace{
    \begin{bmatrix}
        1 \\
        1 \\
        \vdots \\
        1
    \end{bmatrix}}_{\sqrt{N} \ket{+\dots+}} = \frac{\|u\|}{\sqrt{N}} \braket{\phi_u|+\dots+}
\end{equation}

As the velocity field is real, the inner product is real-valued, and its measurement via a Hadamard test enables efficient estimation of the spatial mean.

For error analysis, we compute the single-shot standard deviation of the (noiseless) measurement outcome for $\langle X_a \rangle$. 
Since a projective measurement of $X_a$ yields outcomes $\pm 1$, one can, in order to compute the measurement variance, view it as a Bernoulli random variable; equivalently, using $\langle X_a^2 \rangle = 1$, the variance is
\[
\Delta \langle X_a \rangle = \sqrt{1 - \langle X_a \rangle^2}.
\]
Hence,
\begin{equation}
\begin{split}
    \Delta \expectation{u} 
    &= \frac{||u||}{\sqrt{N}} \Delta \expectation{X_a} = \frac{||u||}{\sqrt{N}} \sqrt{1-\expectation{X_a}^2} = \frac{||u||}{\sqrt{N}} \sqrt{1-\frac{\expectation{u}^2 N}{||u||^2}} = \sqrt{\frac{||u||^2}{N}-\expectation{u}^2}.\label{eq:Bernoulli}
\end{split}
\end{equation}

The number of shots required to achieve an additive precision $\varepsilon$ in $\expectation{u}$ using the Hadamard test scales as $O(1/\varepsilon^2)$. Several alternatives to the Hadamard test can be considered. Amplitude estimation~\cite{brassard2000quantum} provides the theoretically optimal scaling of $O(1/\varepsilon)$ in the number of ansatz circuit calls (corresponding to circuit depth, equivalent to total QPU time). However, the deep and coherent circuits required for amplitude estimation are not yet practical on near-term quantum hardware. Alternatively, direct projective measurements on the state $\ket{+}^{\otimes n}$ yield an efficient estimator for the magnitude of the amplitude. This approach achieves the same $O(1/\varepsilon^2)$ scaling in the number of ansatz calls as the Hadamard test, but with lower circuit depth. It is, however, insensitive  to the sign of $\langle u \rangle$.

The additive statistical uncertainty of estimating $\langle u \rangle$ scales as $\Delta \langle u \rangle \sim \|u\|/\sqrt{N}$, which follows directly from Equation~\eqref{eq:Bernoulli}. In relative terms, this corresponds to an effective signal-to-noise factor $\|u\|^{-1}$, i.e. smaller velocity norms lead to reduced relative precision. In classical fluid dynamics, $\|u\|$ is related to the total kinetic energy in the system (which is proportional to $\|u\|^2$) and its magnitude can vary widely depending on the physical setup. It is determined by the interplay of initial conditions, boundary conditions, dissipation, and external forcing. In decaying flows without external forcing and with dissipative or zero boundary conditions, the norm generally decreases over time due to viscous effects. In contrast, in sustained or driven flows, it may fluctuate around a statistically steady value or even grow temporarily if sufficient energy is injected. Importantly, $\|u\|$ is not fixed by the underlying dynamics alone: in dimensionless formulations, it sets the energy scale but does not affect the qualitative behavior of the solution. As such, the norm can often be rescaled or chosen to balance physical relevance and statistical precision. While small norms can amplify uncertainty in the quantum measurements, they are not fixed by the structure of the equations alone and can be adjusted in dimensionless settings.

\subsection{Estimating \texorpdfstring{$\expectation{u^3}$}{<u³>}}\label{sec:observable_sumk3}
The expectation $\expectation{u^3}=\frac{1}{N}\sum_i u_i^3$ is estimated by running two separate circuits in parallel and measuring the expectation value of a common observable denoted $O_3$. The first circuit  is identical to the Hadamard test used to estimate $\langle u \rangle$, and the second circuit prepares $\ket{\phi_u}$. The state prepared is the following: 
\begin{equation}
\frac{\vert 0 \rangle_a \vert +\dots +\rangle + \vert 1 \rangle_a \vert \phi_u \rangle}{\sqrt{2}} \otimes \vert \phi_u \rangle 
=
\frac{1}{2} \Big(\vert + \rangle_a \cdot \big( \vert +\dots+ \rangle + \vert \phi_u \rangle \big) + \vert - \rangle_a \cdot \big( \vert +\dots+ \rangle - \vert \phi_u \rangle \big) \Big) \otimes \vert \phi_u \rangle,
\end{equation}
where $a$ is the ancilla qubit of the Hadamard test. We choose the observable $O_3$ to be: 
\begin{equation}
\begin{split}
O_3 &= X_a \otimes \sum_i \vert ii \rangle \langle ii \vert = O_3^{(+)}-O_3^{(-)}\\
O_3^{(\pm)} &= \ketbra{\pm}{\pm}_a \otimes \sum_i \ketbra{ii}{ii} .
\end{split}
\end{equation}

The expectation value of $O_3$ yields the scaled sum  $\sum u_i^3$. Note that only the $\ket \pm$ 
branch of the state contributes to the expectation of $O_3^{(\pm)}$, respectively.
\begin{equation} 
\begin{split}
\expectation{O_3^{(\pm)}} &= \frac{1}{4} \tr \bigg( \sum_i \vert ii \rangle \langle ii \vert \cdot \left[ \left( \vert +\dots+ \rangle \pm \vert \phi_u \rangle \right) \cdot \left( \langle +\dots+ \vert \pm \langle \phi_u \vert \right) \otimes \vert \phi_u \rangle \langle \phi_u \vert \right] \bigg)
\\
&= \frac{1}{4} \tr \bigg( \sum_i \vert ii \rangle \langle ii \vert \sum_{j_1, j_2, j_3, j_4} \Big(\frac{1}{2^{n/2}} \pm \frac{u_{j_1}}{\norm{u}} \Big) \Big(\frac{1}{2^{n/2}} \pm \frac{u^*_{j_2}}{\norm{u}}\Big) \frac{u_{j_3} u^*_{j_4}}{\norm{u}^2} \vert j_1 j_3 \rangle \langle j_2 j_4 \vert \bigg) 
\\
&= \frac{1}{4} \sum_i \bigg( \frac{1}{2^{n/2}} \pm \frac{u_i}{\norm u} \bigg)^2 \frac{u_i^2}{\norm u^2} = \frac{1}{4 \norm{u}^2} \sum_i \bigg( \frac{1}{2^n} \pm \frac{2 u_i}{2^{n/2}\norm u} + \frac{u_i^2}{\norm {u}^2}\bigg) u_i^2 
\\
\expectation{O_3}& = \expectation{O_3^{(+)}}-\expectation{O_3^{(-)}} =\frac{1}{4 \norm{u}^2}\sum_i   \frac{  4u_i^3 }{2^{n/2}\norm u} = \frac{1}{\sqrt N \norm{u}^3} \sum_i u_i^3 .
\end{split}
\end{equation}

We note that more advanced approaches, such as those in \cite{holmes2023nonlinear}, are currently impractical on NISQ devices but may become feasible as quantum hardware matures.

\subsection{Estimating \texorpdfstring{$\expectation{u^4}$}{<u⁴>}\label{sec:u4}}
Computing $\expectation{u^4}=\frac{1}{N}\sum_i u_i^4$ follows from the observation that the probability of obtaining identical outcomes when measuring two copies of the state $\ket{\phi_u}$ in  the computational basis is $\sum_i u_i^4$. This holds because the amplitudes of $\ket{\phi_u}$ are real. 
We use a parallel state preparation approach, meaning we prepare the state twice in parallel in partitioned circuits: 
\begin{equation}
    \vert \phi_u \rangle ^{\otimes 2} = \ket{\phi_u} \otimes \ket{\phi_u} .
\end{equation}
By measuring the observable that projects to the subspace where both circuits yield the same output
\begin{equation}
    O_4 = \sum_j \vert jj \rangle \langle jj \vert,
\end{equation}
we obtain 
\begin{align}
    \mathrm{tr} \big( \vert \phi_u \rangle \langle \phi_u \vert ^{\otimes 2} \cdot O_4 \big)= 
   \mathrm{tr} \bigg( \frac{1}{\|u\|^4} \sum_{i_1, i_2, i_3, i_4, j} u_{i_1} u_{i_2} u^*_{i_3} u^*_{i_4} \vert i_1 i_2 \rangle \langle i_3 i_4 \vert \cdot \vert j j \rangle \langle jj \vert \bigg),
\end{align}
where the complex conjugate is only kept here for clarity. Only terms where $i_1=i_2=i_3=i_4=j$ contribute to the sum, and this simplifies to 
\begin{equation}
    \frac{1}{\|u\|^4} \sum_j u^4_j.
\end{equation}

The circuit effectively computes the collision probability when the quantum state is measured in the computational basis. As an alternative to the direct measurement of $O_4$, one could attempt to estimate the same quantity empirically from $M$ computational-basis samples from $|\phi_u\rangle$. The expected number of pairs with identical outcomes among these samples is $\binom{M}{2} \sum_i u_i^4$. However, noise can significantly distort the measurement distribution, leading to substantial bias in this estimation approach.

\subsection{Estimating Shifted Sums\label{sec:shifted_sum}}
The computation of structure functions involves shifted sums of products of velocity components. These quantities can be expressed as expectation values of specific (generally non-diagonal) projectors. To ensure that the observables are Hermitian (and thus physically measurable), we define them as follows:

\begin{align}
    &O^{(r)}_{\text{shifted }1,1} := \frac{1}{2} \sum_j \big( \ketbra{j}{j+r} + \ketbra{j+r}{j} \big) \label{eq:o11} 
    \\
    &O^{(r)}_{\text{shifted }3,1} := \frac{1}{2} \sum_j \big( \ketbra{j,j}{j,j+r} + \ketbra{j,j+r}{j,j} \big) \label{eq:o31} 
    \\
    &O^{(r)}_{\text{shifted }1,3} := \frac{1}{2} \sum_j \big( \ketbra{j,j+r}{j+r,j+r} + \ketbra{j+r,j+r}{j,j+r} \big) \label{eq:o13} 
    \\
    &O^{(r)}_{\text{shifted }2,2} := \sum_j \ketbra{j,j+r}{j,j+r} . \label{eq:o22}
\end{align}

Note that \( O^{(r)}_{\text{shifted }3,1} \) and \( O^{(r)}_{\text{shifted }1,3} \) differ only in the ordering of indices and will yield equivalent quantities with a relabeled shift \( r \).

These observables can be decomposed into Pauli operators and measured efficiently. However, the number of distinct Pauli terms grows exponentially in the general case, and efficient implementations for larger system sizes are left for future work.

The observable in Equation~(\ref{eq:o11}) requires only a single copy of the state $\ket{\phi_u}$, while the remaining expressions in the Equations~(\ref{eq:o31}), (\ref{eq:o13}) and (\ref{eq:o22}) involve two copies $\ket{\phi_u}^{\otimes 2}$, analogous to the setup used for measuring $\sum_i u_i^4$. In each case, the appropriate observable for the chosen shift $r$ is inserted into the measurement circuit, enabling efficient estimation of the respective structure function component. 

\begin{align}
    \frac{1}{\|u\|^2} \sum_i u_i u_{i+r}
    &= \tr \big( \ket{\phi_u} \bra{\phi_u} \cdot O^{(r)}_{\text{shifted }1,1} \big) \label{eq:uiuir}
    \\ 
    \frac{1}{\|u\|^4} \sum_i u_i^3 u_{i+r}
    &= \tr \big( \ket{\phi_u} \bra{\phi_u}^{\otimes 2} \cdot O^{(r)}_{\text{shifted }3,1} \big) \label{eq:ui3uir}
    \\ 
    \frac{1}{\|u\|^4} \sum_i u_i u^3_{i+r}
    &= \tr \big( \ket{\phi_u} \bra{\phi_u}^{\otimes 2} \cdot O^{(r)}_{\text{shifted }1,3} \big) \label{eq:uiuir3}
    \\ 
    \frac{1}{\|u\|^4} \sum_i u_i^2 u_{i+r}^2
    &= \tr \big( \ket{\phi_u} \bra{\phi_u}^{\otimes 2} \cdot O^{(r)}_{\text{shifted }2,2} \big) \label{eq:ui2uir2}
\end{align}

Importantly, once the last three expectation values in Equations~(\ref{eq:ui3uir})-(\ref{eq:ui2uir2}) have been estimated, they must be multiplied by $||u||^4$ before being substituted into Equation~(\ref{eq:4th_sf_lin_comb}). For large values of $||u||$, this scaling significantly amplifies any errors, making the shifted sums highly sensitive to errors, as will be discussed later.

Similar to the estimation  of $\expectation {u^4}$, using samples can be applied to compute the structure functions, but the bias induced by the QPU noise would be significant.

\section{Customized Circuit Design for IBM Heavy-Hex Devices}\label{sec:circuit_design}

\subsection{Parameterized Ansatz Circuit}\label{sec:PAC}
Implementing quantum algorithms on current IBM QPUs involves practical challenges, especially regarding limited qubit connectivity. The most complex circuit in our work arises in the application of the Hadamard test, which requires implementing a controlled-ansatz operation. In this section, we describe how we adapted the brick-wall ansatz to facilitate the implementation of the Hadamard test under these hardware constraints.

\begin{figure}[htbp]
  \centering
  \includegraphics{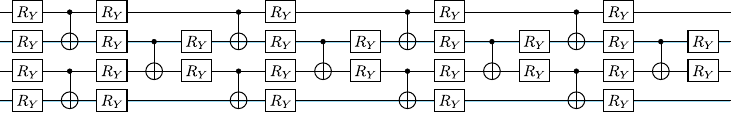}
  \caption{An example of the ansatz architecture featuring a brick wall structure of CNOT and $R_Y$ gates for 4 qubits. An initial layer of $R_Y$ gates is followed by eight layers of CNOT-$R_Y$ blocks.}
  \label{fig:OriginalAnsatz}
\end{figure}

The ansatz circuit needs to fulfill several requirements. It has to be expressive enough to capture a variety of distributions and at the same time it still needs to be trainable. 
This means that a suitable ansatz should contain as many parameters as necessary for a desired accuracy, without redundancy. Furthermore, the ansatz structure can be adjusted regarding the characteristics of the data to be displayed. Since the velocity takes only real values, the ansatz circuit can be constructed using $R_Y$ rotations and CNOT gates for entanglement. This combination is expressive enough to generate a wide range of real-valued quantum states, although due to circuit depth limitations not all possible states may be accessible. We start with the canonical brick-wall ansatz (see Figure~\ref{fig:OriginalAnsatz}) similar to~\cite{Siegl_2025}. 

This study uses IBM's Heron r2 device \texttt{ibm\_fez}, which features a heavy-hex topology~\cite{ibm_quantum_systems}. In this layout, each qubit is connected to two or three others (as can be seen in Figure~\ref{fig:HeavyHex}), limiting the direct applicability of gate operations that require all-to-all connectivity.

\begin{figure}[htbp]
  \centering
  \includegraphics{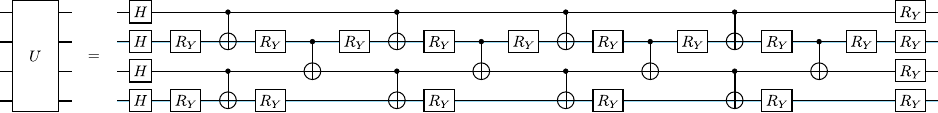}
  \caption{Ansatz architecture, shown here for 4 qubits as an example, incorporating adjustments regarding given hardware requirements. Compared to the original ansatz architecture, it includes an initial layer of Hadamard gates, rotations only on every second qubit within the eight CNOT-$R_Y$ ansatz layers and a final layer of rotations on every qubit.}
  \label{fig:Ansatz}
\end{figure}

The ansatz developed here is designed to remain compatible with these constraints while maintaining sufficient expressivity. To this end, $R_Y$ rotations are limited to odd qubits. This reduction in the ansatz degrees of freedom relaxes the complexity required to implement the control ansatz in the heavy-hex connectivity as we will show in Section ~\ref{sec:mean}. However, this also leads to a reduction in expressivity, which is recovered to a certain extend by adding a final layer of $R_Y$ rotations to all qubits (see Figure \ref{fig:Ansatz}).

Furthermore, a layer of Hadamard gates is included at the beginning of the ansatz circuit. This design choice was made with regard to the later use of a Hadamard test (see Section~\ref{sec:mean}), because this way the ansatz inherently contains both states that are later required for the Hadamard test. 
With all $R_Y$ angles set to zero, the Hadamards alone prepare the uniform reference state $\ket{+}^{\otimes n}$, while activating the $R_Y$ rotations produces the velocity-encoding state. Within the Hadamard test, it is therefore sufficient to control only the single-qubit $R_Y$ rotations. The CNOTs do not need to be controlled, i.e. replaced by gate-intensive Toffolis, since they are part of both states.  
Combining all described characteristics, the circuit features an architecture as illustrated in Figure~\ref{fig:Ansatz}. For illustration, we show the construction in the 4-qubit case, but the ansatz scales straightforwardly to larger system sizes.

A dimensional expressivity analysis (DEA)~\cite{Funcke_2021} was performed to assess the expressivity of the adjusted ansatz (see Appendix~\ref{sec:DEA}). The DEA quantifies how many independent directions in Hilbert space the ansatz can explore, based on the rank of the Jacobian with respect to its parameters. For a real-valued 4-qubit circuit, the maximum attainable rank is 15, whereas the adjusted ansatz presented here reaches a rank of 7, indicating redundant parameters and limited local expressivity. Nevertheless, further numerical tests showed that randomly sampled target states could still be approximated with high accuracy, with final errors on the order of $10^{-11}$. 
Future studies could explore introducing SWAP gates between CNOT-$R_Y$ layers to improve the expressivity of the circuit.

\subsection{Measurement Circuits}

\subsubsection{Measuring \texorpdfstring{$\expectation{u}$}{<u>}}\label{sec:mean}

Standard implementations of the Hadamard test use a single auxiliary qubit that controls the entire ansatz circuit. On a heavy-hex architecture, however, this qubit would need to be routed through all active regions of the circuit, resulting in a large number of SWAP operations and correspondingly deep circuits. To avoid this overhead, we adopt a distributed-control strategy. In the general case, we place an auxiliary qubit next to every second circuit qubit and entangle all auxiliary qubits into a cat state $\ket{00\dots0} + \ket{11\dots1}$. This allows the control to be applied locally within each subregion of the circuit, while the cat state ensures global coherence of the controlled operation. For a 4-qubit ansatz, this reduces to two auxiliary qubits $a_1$ and $a_3$. 

The overhead for preparing the cat state depends only on the topological distance between the auxiliary qubits. In particular, creating a cat state over a path of length $d$ requires $O(d)$ two-qubit-operations, i.e. linear overhead in the separation (sub-linear depth can be achieved using mid-circuit measurements and classical feed-forward \cite{siddardha2025shallow}). In practice, this leads to more than three-fold reduction in depth compared to routing a single global control qubit with standard transpilers from Qiskit~\cite{qiskit2024}  and BQSKit~\cite{BQSKIT}.

\begin{figure}[ht]
    \centering
    \includegraphics[width=0.4\linewidth]{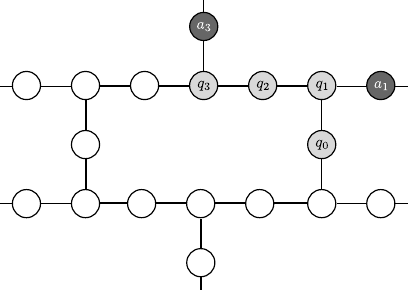}
    \captionof{figure}{Example of qubit positioning on IBM QPU connectivity. The circuit qubits are shown in light gray, the adjacent auxiliary qubits are shown in dark gray.}
    \label{fig:HeavyHex}
\end{figure}

\begin{figure}[htbp]
  \centering
  \includegraphics{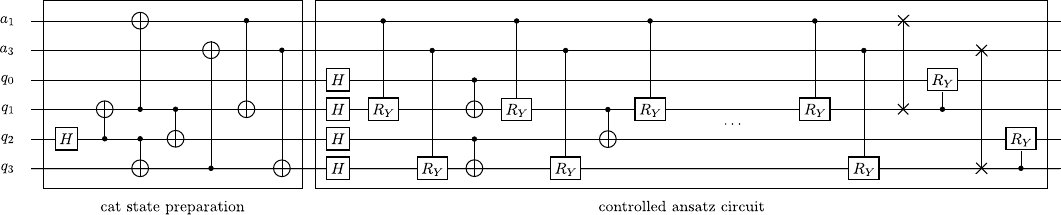}
  \caption{Hadamard test implementation. Two auxiliary qubits $a_1,a_3$, prepared in the cat state, are used to control exclusively the $R_y$ rotations in the ansatz circuit. Throughout most of the circuit, the controlled rotations act on   $q_1$ and $q_3$. For controlling the ansatz's last $R_y$ layer, two SWAP gates  are performed: $a_1\leftrightarrow q_1, ~a_3\leftrightarrow q_3$, allowing the ancillas $a_1,a_3$ to apply ctrl-$R_y$ on qubits $q_0,q_2$, respectively. Finally, $\expectation{X_{q_1}X_{q_3}}$ is estimated (the ancilla qubits are in $q_1,q_3$ after the SWAPs).}
  \label{fig:adjusted_hadamard}
\end{figure}

An example of a qubit layout for 4 circuit qubits is shown in Figure~\ref{fig:HeavyHex}, where the auxiliary and ansatz qubits are placed in a way to minimize qubit swapping. As depicted in Figure~\ref{fig:adjusted_hadamard}, the cat state preparation takes 4 CNOT-layers for 4 circuit qubits and 2 auxiliaries. However, this schemes also generalizes to larger encodings provided the auxiliary qubits are placed adjacent to their respective controlled circuit qubits. Under such layouts the SWAP overhead remains local and the method scales favorably. Consequently, the Hadamard test observable $X_a$ is replaced by $\bigotimes_{a\in cat}X_a$. Since the cat qubits are placed adjacent to the active ansatz qubits, they can be used as control qubits for the controlled $R_Y$ gates without any additional SWAP gates. Note that controlling different subblocks of the ansatz via distinct auxiliary qubits (entangled in a cat state) is not algebraically identical to a single global multi-control, but for our construction it yields the same interference needed in the Hadamard test while avoiding expensive multi-control gates.

Crucially, this approach leverages the structure of the ansatz described in Section~\ref{sec:PAC}: with all rotation angles set to zero, the ansatz produces the uniform state $\ket{+}^{\otimes n}$. By promoting each $R_Y$ gate to a controlled $R_Y$ gate (with control from the cat state), we implement a clean switching mechanism: at the $\ket{00}$ branch of the ancilla qubits superposition, the controlled rotations remain inactive, and only the Hadamard and CNOT gates are applied, effectively preparing the state $\ket{+}^{\otimes n}$. In contrast, at the  $\ket{11}$ branch, the controlled rotations are activated, preparing the velocity-encoding state. This controlled switching mechanism is illustrated schematically in Figure~\ref{fig:adjusted_hadamard}.

A final BQSKIT optimization pass is applied to the circuit in Fig. \ref{fig:adjusted_hadamard}, further reducing the overall circuit depth. A summary of key properties of this adjusted Hadamard test circuit is provided in Table~\ref{tab:hadamard_test_circuit_properties}. This construction enables the Hadamard test to be realized under strict hardware constraints, with minimal qubit swapping and a reduced gate count.

\begin{table}[htbp]
    \centering
    \begin{tabular}{lr}
        \toprule
            Circuit Property & Value \\
        \midrule
            Number of qubits & 6 \\
            Number of two-qubit gates & 39 \\
            Number of two-qubit layers & 24 \\
        \bottomrule
    \end{tabular}
    \captionof{table}{Key properties of the adjusted Hadamard test quantum circuit.}
    \label{tab:hadamard_test_circuit_properties}
\end{table}

\subsubsection{Measuring \texorpdfstring{$\expectation{u^3}$}{<u³>}} 
\begin{figure}[htbp]
  \centering
  \includegraphics{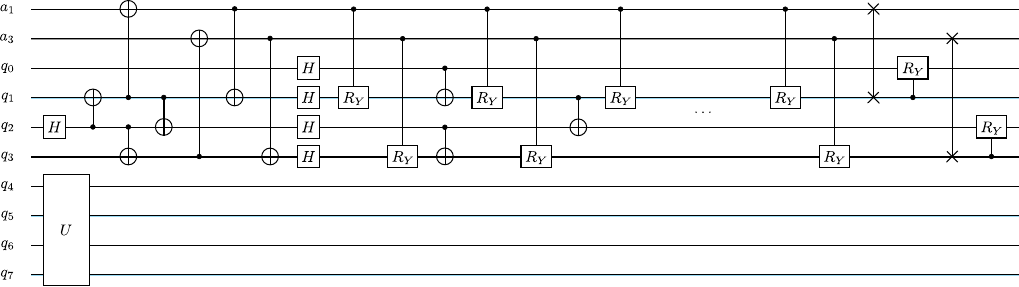}
  \caption{Circuit for measuring the scaled sum of the amplitudes cubed consisting of the circuit for measuring the scaled sum of the amplitudes and the ansatz in parallel.} \label{fig:circuit_k3}
\end{figure}

 As described in Section \ref{sec:observable_sumk3}, the sum $\sum_i u_i^3$ requires running two circuits  in parallel: the Hadamard test and the ansatz (see Figure~\ref{fig:circuit_k3}). 

Recall the observable $O_3$: 
\begin{equation}
\begin{split}
O_3 &= X_a \otimes \sum_i \vert ii \rangle \langle ii \vert = O_3^{(+)}-O_3^{(-)}\\
O_3^{(\pm)} &= \ketbra{\pm}{\pm}_a \otimes \sum_i \ketbra{ii}{ii}.
\end{split}
\end{equation}
Here, the projector $\sum_i \ketbra{ii}{ii}$ acts on the two copies of the system register: $q_0\dots q_3$ and $q_4\dots q_7$. Measuring the expectation value of this observable yields the desired quantity $\sum_i u_i^3$ (up to a global factor), since it projects onto the diagonal elements of the reduced density matrix, effectively isolating the cube of each amplitude. 

The key resource metrics are summarized in Table~\ref{tab:sum3_circuit_properties}. Notably, despite the increased qubit count, the overall circuit depth remains unchanged due to the parallel structure of the ansatz preparation.

\begin{table}[htbp]
\centering
\begin{tabular}{lr}
\toprule
Circuit Property & Value \\
\midrule
Number of qubits    & 10 \\
Number of two-qubit gates & 51 \\
Number of two-qubit layers & 24 \\
\bottomrule
\end{tabular}
\caption{Key properties of the quantum circuit for measuring the sum of the amplitudes cubed.}
\label{tab:sum3_circuit_properties}
\end{table}

\subsubsection{Measuring \texorpdfstring{$\expectation{u^4}$}{<u⁴>} and Shifted Sums}
Computing $\expectation{u^4}$ and shifted sums is done by running two   state preparations in parallel with a common observable as described in Sections \ref{sec:u4} and \ref{sec:shifted_sum} to form  the tensor product state $\ket{\phi_u}^{\otimes 2}$.

In particular, the sum of amplitudes to the power of four, $\sum_i u_i^4$, corresponds to the probability of observing identical outcomes on both state copies. This is realized by measuring the observable

\begin{equation}
    O_4=\sum_j \ketbra{jj}{jj},
\end{equation}

whose expectation value selects diagonal elements of the product density matrix and thus isolates the quartic terms.

\begin{figure}[htbp]
  \centering
  \begin{minipage}[t]{0.48\textwidth}
    \centering
    \includegraphics[width=0.2\linewidth]{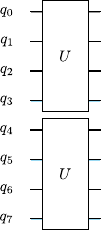}
  \end{minipage}
  \hfill
  \begin{minipage}[t]{0.48\textwidth}
      \centering
      \vspace*{-2.05cm}
      \begin{tabular}{lr}
        \toprule
        Circuit Property & Value \\
        \midrule
        Number of qubits    & 8 \\
        Number of two-qubit gates & 24 \\
        Number of two-qubit layers & 8 \\
        \bottomrule
    \end{tabular}
  \end{minipage}
  \caption{Circuit layout and corresponding key properties for measuring the sum of the amplitudes to the power of four.}\label{fig:circuit_k4}
\end{figure}

A schematic overview of the circuit implementing this measurement, along with its key properties, is provided in Figure~\ref{fig:circuit_k4}. The implementation requires only the ansatz depth and twice the number of ansatz qubits. 

All sums of the form \(\sum_i u_i^m u_{i+r}^\ell\), including the special case \(\sum_i u_i^4\), can be measured  a shared circuit structure, where the ansatz state \(\ket{\phi_u}\) is prepared once (for \(m+\ell=2\)) or in two copies \(\ket{\phi_u}^{\otimes 2}\) (for \(m+\ell=4\)). Crucially, the measurement observable is the only component that needs to be modified to extract different moment combinations. For instance, measuring the observable \( O^{(r)}_{\text{shifted }2,2} = \sum_j \ketbra{j,j+r}{j,j+r} \) defined in Equation~(\ref{eq:o22}) yields the quantity $\sum_i u_i^2 u_{i+r}^2$ up to normalization. Similarly, the non-diagonal operators \( O_{\text{shifted 3,1}}^{(r)} \) and \( O^{(r)}_{\text{shifted }1,3} \) defined in Equations~(\ref{eq:o31}) and (\ref{eq:o13}) allow access to asymmetric combinations such as $\sum_i u_i^3 u_{i+r}$ or $\sum_i u_i u_{i+r}^3$. This enables us to reuse the QPU measurement outcomes, thus significantly reducing experimental overhead.

\section{Hardware Results}\label{sec:hardware_results}
To evaluate the feasibility of extracting statistical properties of velocity fields on current quantum hardware, we deployed our quantum circuits on IBM’s superconducting quantum device Heron r2 \texttt{ibm\_fez}. This device features a 156-qubit heavy-hexagonal lattice and includes TLS noise mitigation for improved coherence and stability. 

To reduce the QPU usage and the total number of jobs, we parallelized the measurement of several observables and time steps by combining them into a single circuit execution. Specifically, the sum $\sum_i{u_i^4}$ and the shifted sums were measured simultaneously for two distinct spatial distributions by embedding two independent 8-qubits circuits in parallel on a 16-qubit system, allowing to evaluate all 66 observables required for both spatial distributions within one job. 
As an error suppression and error mitigation tool, we used the QESEM software~\cite{QESEM}. The target precision per observable was set between 0.1 and 0.05, and each job (i.e., a single submission to the QPU that may contain measurements of multiple observables bundled into one circuit) was allocated a runtime limit between 12 minutes and one hour, depending on the complexity of the circuit. Despite the conservative time limits, the actual QPU time usage per job remained below 3 minutes, yielding a total QPU usage of not more than 40 minutes for all results shown in this work. 

In the following sections, we present and analyze the results obtained from executing our circuits on real quantum hardware, including the computation of spatial velocity moments and preliminary structure functions.

\subsection{Sine Signal Test Case}\label{sec:sine_test_case}

We benchmark the statistical moment estimation circuits with respect to QPU runtime and precision, using a simple discrete test signal based on a sine function. It is defined on \(2^n\) equidistant points in the interval \([0,2\pi]\) for two resolutions, on 4 qubits yielding 16 lattice points and on 8 qubits yielding 256 lattice points:
\begin{equation}
    u_i = \frac{\mathrm{sin} \left( \frac{2 \pi i}{2^n} \right) + 1}{\sqrt{\sum_{j=0}^{2^n-1} \big[ \mathrm{sin} \left( \frac{2 \pi j}{2^n} \right) + 1 \big]^2}}, \quad i=0,\cdots, 2^n-1.
\end{equation}
The constant shift ensures that the mean value is nonzero and the normalization eliminates scaling effects, yielding a unit-norm vector \(\|u\|=1\). The ansatz is then classically optimized to prepare the state:
\begin{equation}
    \ket{\phi_u}= \sum_j u_j\ket{j} .
\end{equation}

\begin{figure}[ht!]
    \begin{center}
        \resizebox{.8\textwidth}{!}{%
            \includegraphics[height=3cm]{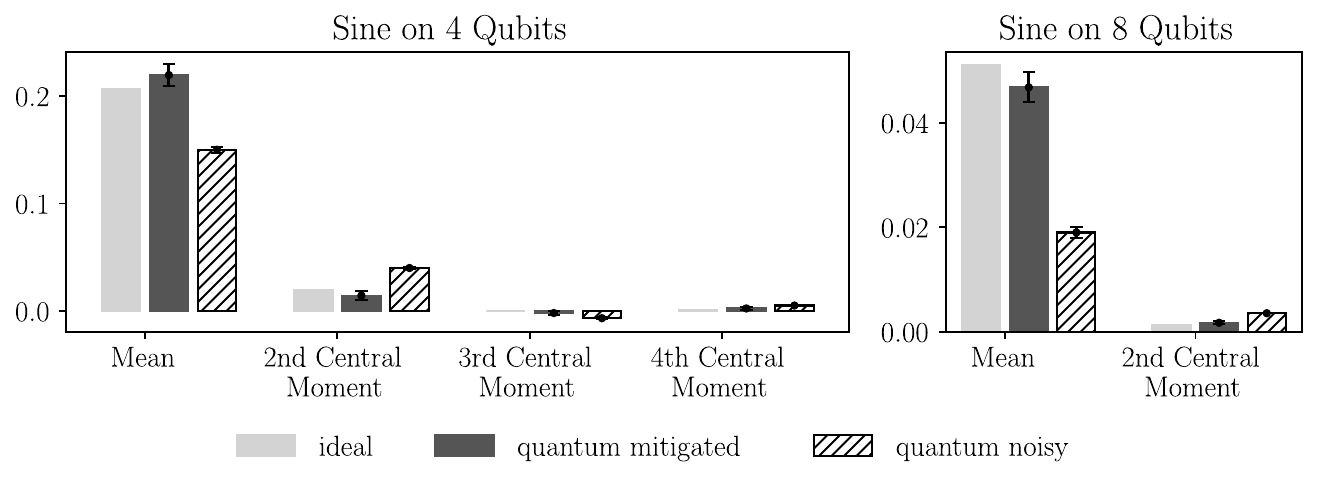}
        }

        \vspace{1em}

        \begin{tabular}{crrrrrr}
            \toprule
             & \multicolumn{4}{c}{Sine on 4 Qubits} & \multicolumn{2}{c}{Sine on 8 Qubits} \\
            \midrule
             & Mean & \makecell{2nd Central\\Moment} & \makecell{3rd Central\\Moment} & \makecell{4th Central\\Moment} & Mean & \makecell{2nd Central\\Moment} \\
            \midrule
            \makecell[c]{Classical Reference\\$x_{\text{ref}}$}   & 0.2041 & 0.0208 & 0 & 0.0007 & 0.0510 & 0.0013 \\
            \midrule
            \makecell[c]{Statevector\\Simulation} & 0.2063 & 0.0199 & 0 & 0.0006 & 0.0511 & 0.0013 \\
            \midrule
            Noisy Value & \makecell[r]{0.1501\\$\pm$ 0.0028} & \makecell[r]{0.0399\\$\pm$ 0.0008} & \makecell[r]{-0.0067\\$\pm$ 0.0007} & \makecell[r]{0.0051\\$\pm$ 0.0005} & \makecell[r]{0.0190\\$\pm$ 0.0011} & \makecell[r]{0.0035\\$\pm$ 0.0001} \\
            \midrule
            \makecell[c]{Mitigated Value\\$\hat{x} \pm \Delta x$} & \makecell[r]{0.2195\\$\pm$ 0.0099} & \makecell[r]{0.0143\\$\pm$ 0.0043} & \makecell[r]{-0.0019\\$\pm$ 0.0015} & \makecell[r]{0.0024\\$\pm$ 0.0010} & \makecell[r]{0.0468\\$\pm$ 0.0029} & \makecell[r]{0.0017\\$\pm$ 0.0003} \\
            \midrule
            \makecell[c]{Normalized Deviation\\$N_\sigma = \frac{|\hat{x} - x_{\text{ref}}|}{\Delta x}$} & 1.34 & 1.3 & 1.28 & 1.66 & 1.48 & 1.33\\
            \bottomrule
        \end{tabular}
    \end{center}
    \caption{Comparison of quantum-computed central moments of the sine signal test case with their ideal reference values. Results are shown for four cases: classical reference values, ideal statevector simulation, noisy QPU measurements (without error mitigation), and mitigated QPU measurements (with error mitigation).}
    \label{fig:central_moments_sine}
\end{figure}

We compare quantum-computed central moments with their classically expected ideal values for sine signals encoded on $n=4$ and $n=8$ qubits, corresponding to $2^4$ and $2^8$ spatial points, respectively. The results are shown in Figure~\ref{fig:central_moments_sine}. To quantify discrepancies, we compute the normalized deviation between the measured values and ideal values as
\begin{equation}
    N_\sigma = \frac{|\hat{x}-x_{\text{ref}}|}{\Delta x},
\end{equation}
commonly referred to as the z-score or $N_\sigma$ deviation. Here, $\hat{x}$ denotes the quantum estimate, $x_{\text{ref}}$ the ideal classical value, and $\Delta x$ its uncertainty obtained by error propagating the QESEM-reported error bars. For all moments, the normalized deviation remains below 2, indicating statistical consistency with the reference. This threshold corresponds to approximately two standard deviations, within which values are statistically expected to lie with high probability under normal measurement noise. 

\begin{figure}[!htbp]
    \centering
        \includegraphics[width=.7\linewidth]{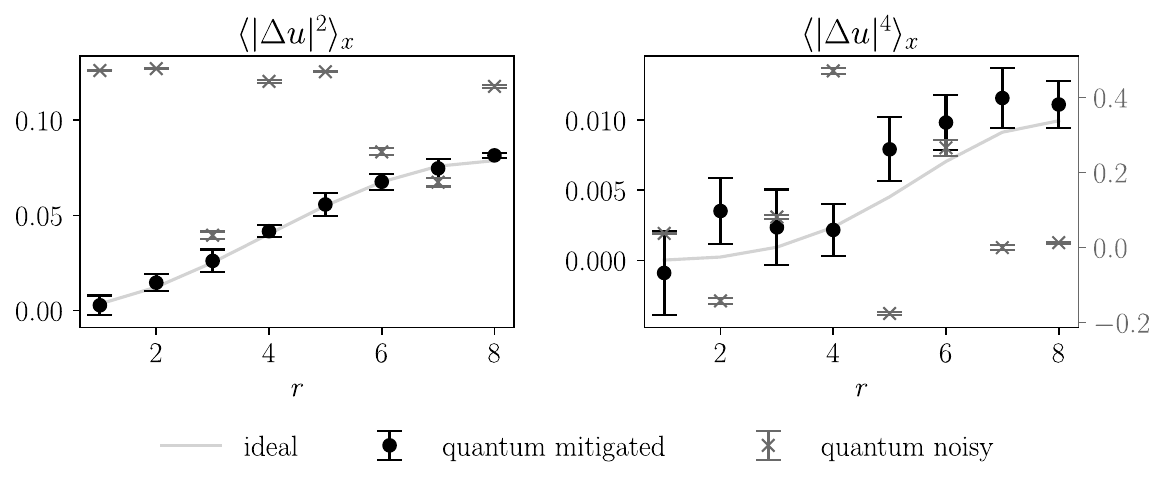}
    \caption{Spatially averaged second-order (left) and fourth-order (right) structure functions for the sine signal test case. The left y-scale in the fourth-order plot refers to the ideal and the mitigated quantum values, whereas the right y-scale refers to the noisy quantum results. Without error mitigation, the results deviate significantly from the expected behavior. \label{fig:sf_sine}}
\end{figure}

In addition to the central moments, the spatially averaged second-order and fourth-order structure functions for the sine signal velocity field are depicted in Figure~\ref{fig:sf_sine}. As expected, the velocity differences increase with the spatial separation length $r$ due to the smooth periodic nature of the sine function and the large wavelength of the signal, which causes velocity differences to accumulate gradually over distance. 

For the second-order structure function (left in Figure~\ref{fig:sf_sine}), the quantum results are in excellent agreement with the classical baseline. The values almost always overlap or fall within the error bars, demonstrating the robustness of the quantum computation for this observable. 

In the fourth-order case (right in Figure~\ref{fig:sf_sine}), the error bars become larger, reflecting increased statistical uncertainty. Nevertheless, the mitigated data are still statistically consistent with the ideal values within these uncertainties. This behavior mainly arises from the normalization of the velocity vectors, which causes the absolute values of the fourth-order structure function to be very small. Consequently, even small statistical errors in the quantities measured on the QPU lead to noticeable relative deviations. When rescaled to non-normalized units, these differences grow accordingly, which also explains the increased uncertainty, and the large biases of the noisy values.

\subsection{Burgers' Turbulence}\label{sec:burgers_turbulence}

Our main result is measuring the statistical moments of snapshots in a realistic evolution of a velocity field governed by the 1D Burgers’ equation. We performed the evolution and optimized parameters on classical hardware, deferring the quantum implementation to future work.

The equation is solved using a pseudo-spectral method on a spatial grid of $N=16$ points with periodic boundary conditions. The spatial domain extends from $x_b=0$ to $x_e=10 \times 2\pi$, and the grid spacing is defined as:
\begin{equation}
    dx=\frac{x_e - x_b}{N} .
\end{equation}
A time step of $dt=0.01$ is used, and the simulation is run for a total of 54,000 time steps. The viscosity is set to $\nu=0.1$.

The initial velocity field is set to zero, and a forcing term is applied at each time step to drive the system. The chosen forcing function is designed to generate shock waves in the system and is characterized by being white in time and self-similar in space. Specifically, the forcing term is defined in Fourier space as:
\begin{equation}
    f_k(t) = D_0 \vert k \vert ^{\beta} \xi_k(t)
\end{equation}
where $D_0=0.5$, $\beta = -1$, and $\xi_k(t)$ is a complex-valued Gaussian noise term with independent real and imaginary parts, each drawn from a normal distribution with variance $1/2$. This formulation ensures that the forcing exhibits the desired statistical properties, with energy injection occurring primarily at larger scales.

\begin{figure}[ht!]
    \centering
    \includegraphics[width=0.97\linewidth]{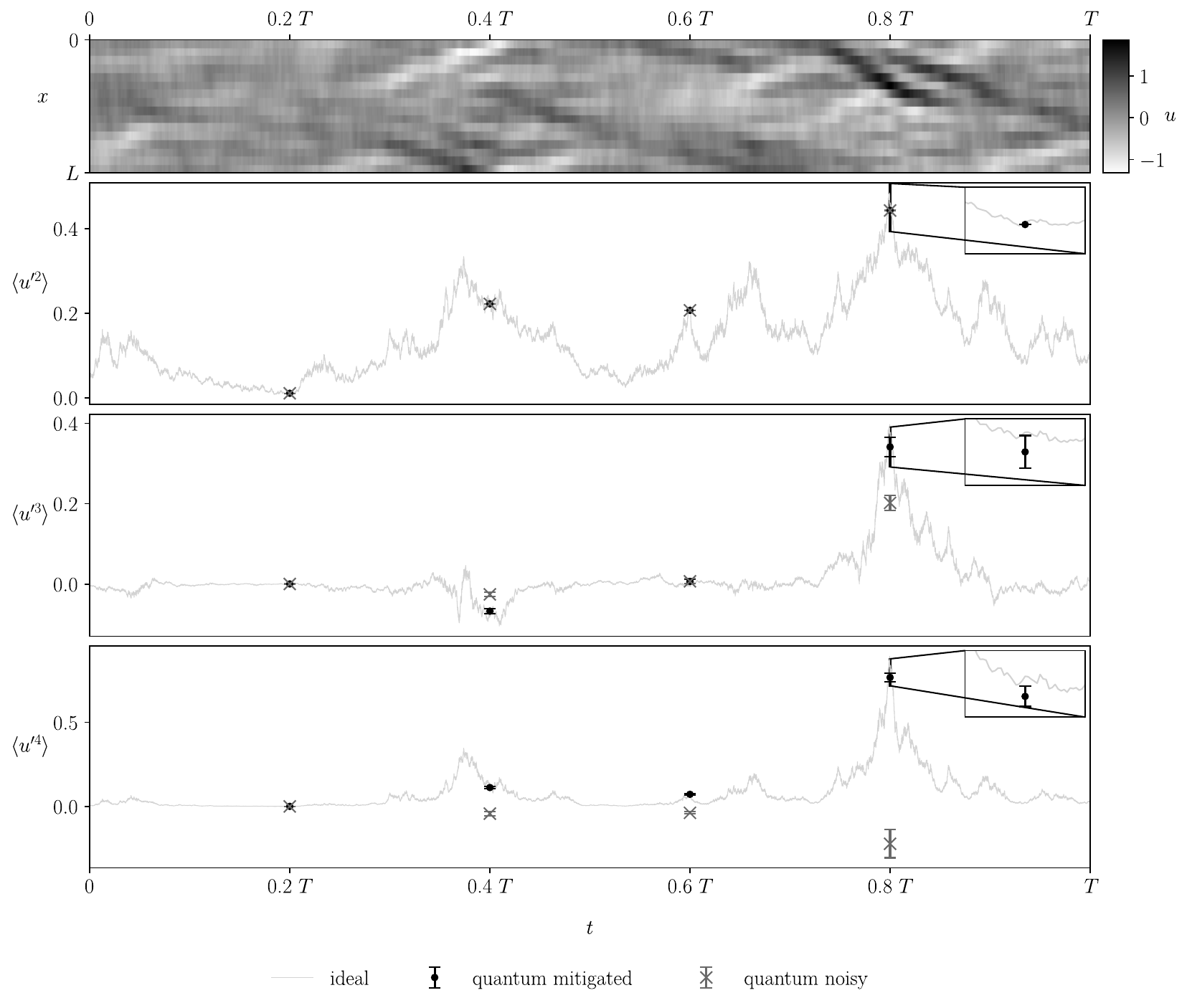}
    \caption{Spatiotemporal evolution of the velocity field (top panel) and temporal evolution of the second, third and fourth central moment (second, third and fourth panel) in the forced Burgers' simulation. Sharp gradients in the velocity field indicate shock formation, coinciding with extrema in the statistical moments. Classical reference values are shown as gray lines. Error-mitigated quantum results are depicted as black dots with error bars, while unmitigated (noisy) results are shown as dark gray crosses with error bars, at selected time steps. All mitigated quantum estimates remain within approximately two standard deviations, indicating statistical consistency. }
    \label{fig:moments_over_time}
\end{figure}

Figure~\ref{fig:moments_over_time} shows the spatiotemporal evolution of the velocity field governed by the forced Burgers' equation (top panel), as well as the evolution of its statistical moments over time (bottom panels). Classical reference curves are shown continuously, while quantum-computed results are presented at selected points in time.
In the top panel, shock-like structures emerge as the system evolves, though the coarse resolution limits the complexity of the resulting patterns. To examine different dynamical regimes, four representative time points are selected at $0.2T, 0.4T, 0.6T$ and $0.8T$, where $T$ denotes the total simulation time. This selection includes calm phases, distinct shock formation, and transitional stages between quiescence and turbulence.

Throughout the evolution, the error-mitigated quantum estimates closely follow the classically computed reference values. Their deviations remain within two standard deviations, indicating statistical consistency. 
At time $0.8T$, where a pronounced shock structure leads to more extreme statistical features, the second, third, and fourth central moments increase substantially. While the mitigated results still capture the overall trends, minor deviations and slightly larger uncertainties become visible in this regime. The unmitigated (noisy) values, by contrast, show stronger deviations, particularly in the higher-order moments and during dynamically active phases (e.g., $0.4T$ and $0.8T$) where a large bias of the noisy results can be observed. 

\begin{figure}[ht!]
    \centering
    \begin{subfigure}{\textwidth}
        \includegraphics[width=\textwidth]{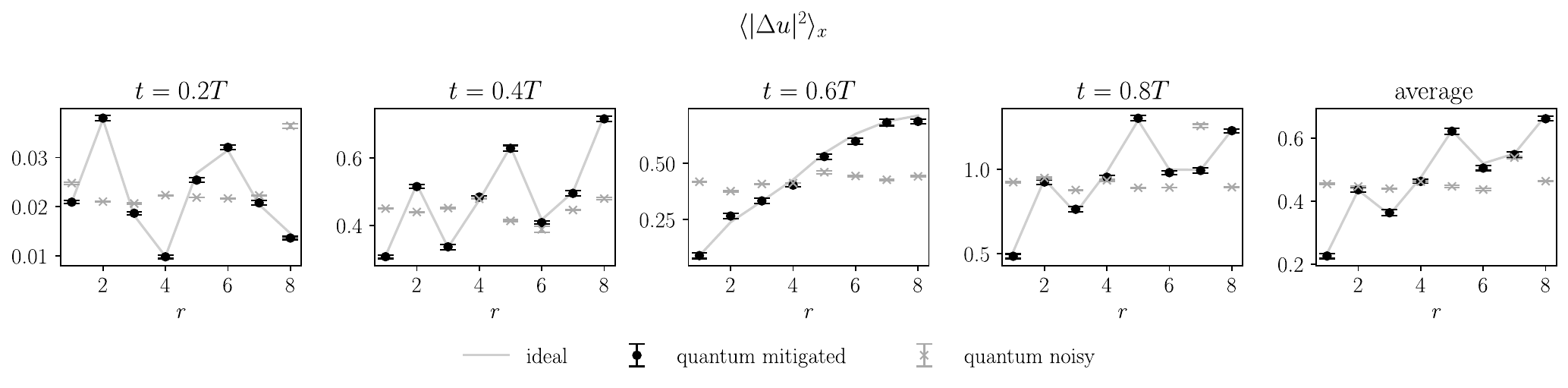}    
        \caption{Spatially averaged second-order structure function $S_2(r,t)$.}\label{fig:second_order_structure_function_burgers}
    \end{subfigure}
    \vskip\baselineskip
    \begin{subfigure}{\textwidth}
        \includegraphics[width=\textwidth]{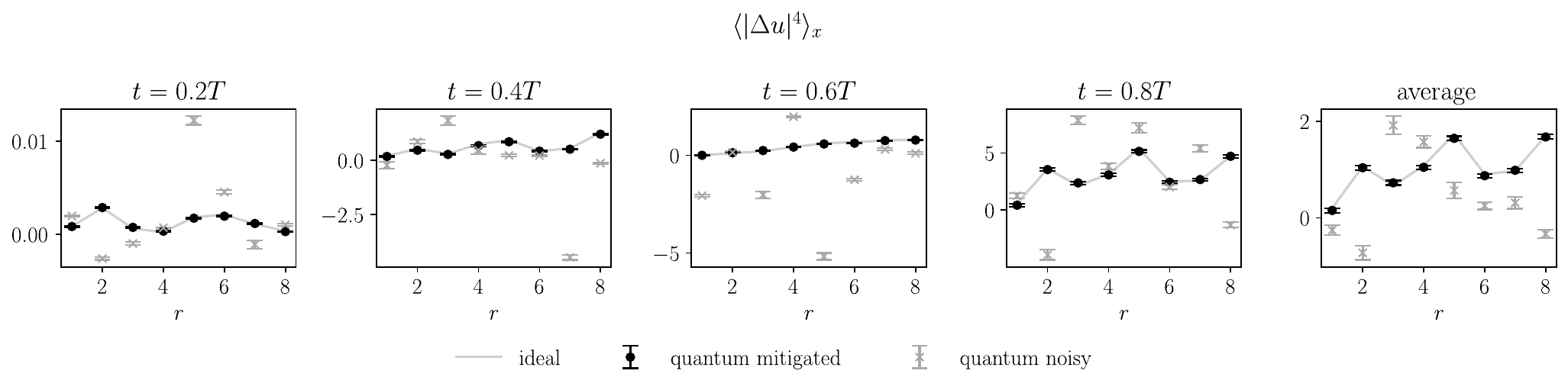}
        \caption{Spatially averaged fourth-order structure function $S_2(r,t)$.}\label{fig:fourth_order_strcuture_function_burgers}
    \end{subfigure}
    \caption{Spatially averaged structure functions of second and fourth order at different simulation time steps $t$. The mitigated quantum results closely match the ideal values. Without error mitigation, the results deviate significantly from the expected behavior as the calculation involves a scaling of $||u||^2$ or $||u||^4$ of the estimated expectation values, amplifying errors. The norms of the spatial velocity distributions at the given time steps are $\|u(t=0.2T)\|=0.42$, $\|u(t=0.4T)\|=1.88$, $\|u(t=0.6T)\|=1.82$, and $\|u(t=0.8T)\|=2.66$ (dimensionless, due to nondimensionalization of the governing equations).}
\end{figure}

Figures \ref{fig:second_order_structure_function_burgers} and \ref{fig:fourth_order_strcuture_function_burgers} show the second-order and fourth-order structure functions, respectively, computed at times $0.2T$, $0.4T$, $0.6T$, and $0.8T$, along with the average over these four snapshots.  In fully developed turbulence, structure functions, i.e. statistical moments of velocity differences at varying spatial separations, are expected to exhibit characteristic scaling behavior. According to the classical Kolmogorov–Obukhov theory from 1941~\cite{kolmogorov1941, obukhov1941}, the $k$-th order structure function scales with the spatial offset $r$ as $r^{k/3}$. Later refinements, including intermittency corrections by She and Leveque~\cite{she1994}, modified this to include deviations from simple scaling, especially at higher orders~\cite{Birnir_2014}. These scaling laws have become a central benchmark in the analysis of turbulent flows, including Burgers' turbulence, where shock-dominated dynamics lead to distinct deviations from Gaussian statistics.

Due to the very coarse spatial resolution of only 16 grid points, these results cannot be used to extract scaling laws or study inertial-range behavior, which would require both finer spatial discretization and averaging over a large ensemble of velocity fields. Nevertheless, the structure functions do reveal physically meaningful features. At certain time steps, such as $t=0.6T$, the structure functions increase nearly monotonically with spatial separation $r$, which indicates the presence of long-range velocity correlations and coherent flow structures. At other times, oscillatory or non-monotonic patterns emerge (e.g., $t=0.2T$, $0.4T$ or $0.8T$), reflecting characteristic length scales or competing gradients in the velocity field. The time-averaged structure functions exhibit smoother behavior and suggest a saturation at larger $r$, hinting at the existence of dominant structure sizes even within this coarse representation. 

While the classical results  (light gray lines) clearly show these features, they are not visible at all in the noisy results (dark gray crosses with error bars) as the noise induced bias effectively washes them out. Importantly, the error-mitigated quantum results (black points with error bars) closely match the classical reference solutions. In nearly all cases, the quantum estimates lie well within two standard deviations of the classical values, demonstrating the statistical reliability of the method. While the spatial resolution limits the physical interpretability of fine-scale behavior, the results provide a proof of principle that structure function analysis is feasible on near-term quantum hardware and yields trustworthy outcomes, provided that suitable error mitigation techniques are applied.

\section{Discussion and Outlook}\label{sec:discussion}

A central bottleneck in Quantum Computational Fluid Dynamics is the readout of physically meaningful quantities at scale: full state tomography is infeasible, and much of the literature emphasizes the quantum algorithm itself, while the extraction of interpretable outputs is often left under-specified. We address this by presenting measurement schemes that allow to recover spatial statistics from amplitude-encoded velocity fields without full state tomography. This demonstrates a proof-of-concept QCFD pipeline with a practical readout layer that converts compact amplitude-encoded states into actionable statistics while reducing  post-processing overhead. 

Our results show that nontrivial statistical quantities, such as central moments and second- and fourth-order structure functions, can be obtained reliably on current quantum devices. The error mitigated quantum estimates are consistent with classical reference values, typically agreeing within one to two standard deviations. Further improvements in precision are expected to be bounded mainly by current quantum device limitations and the QPU time devoted to error mitigation.

A natural next step is to extend the method to larger quantum systems in order to resolve finer spatial features and probe statistical signatures such as scaling laws or intermittency effects in turbulent flows. However, the scalability of the current approach is fundamentally tied to the encoding strategy, the expressivity and depth of the ansatz circuit and the structure of the measured observables. All measurement schemes rely on normalization with respect to the velocity field norm $\|u\|$, whose scaling with system size depends on the physical setup and normalization conventions. While the expectation value used to estimate the mean scales with $\|u\|/\sqrt{N}$, higher-order moments introduce stronger norm dependencies, such as ${N}^{-1/2}\|u\|^{-3}$ or $\|u\|^{-4}$, making them more sensitive to both noise and fluctuations in $\|u\|$. For a fixed target precision, these dependencies increase the required estimator accuracy and, consequently, the QPU time.  Future work will therefore need to carefully balance physical fidelity, norm control, and measurement overhead to ensure scalability beyond proof-of-concept implementations.

\section*{Acknowledgments}
We sincerely thank Greta Sophie Reese and Takis Angelides for the valuable discussions and constructive suggestions, which significantly contributed to the progress of this work. We are also grateful to Stefan Kühn for insightful advice regarding the dimensional expressivity analysis, and to Eyal Bairey and Omri Golan for their support and helpful input throughout the project. 

This work is supported with funds from the Ministry of Science, Research and Culture of the State of Brandenburg within the Center for Quantum Technology and Applications (CQTA).

\begin{figure}[ht!]
\centering
    \includegraphics[width=0.2\linewidth]{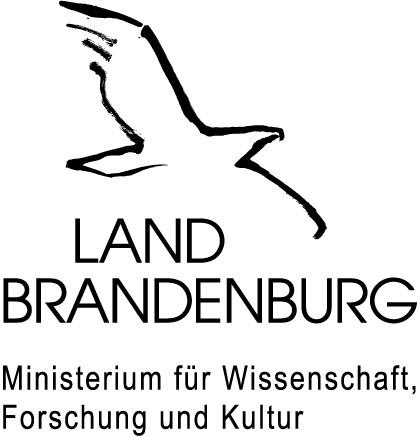}
\end{figure}

\printbibliography

\newpage
\begin{appendices}

Appendix~\ref{sec:Appendix_Hardware} provides all obtained hardware measurement results, complementing the data shown in the main text. Appendix~\ref{sec:QESEM} describes the QESEM error suppression and mitigation framework used in this work. Finally, Appendix~\ref{sec:DEA} presents the dimensional expressivity analysis of the ansatz circuits.

\section{Hardware Measurement Results}\label{sec:Appendix_Hardware}

\begin{figure}[ht!]
    \centering
    \begin{subfigure}[b]{.45\textwidth}
        \centering
        \includegraphics[width=\linewidth]{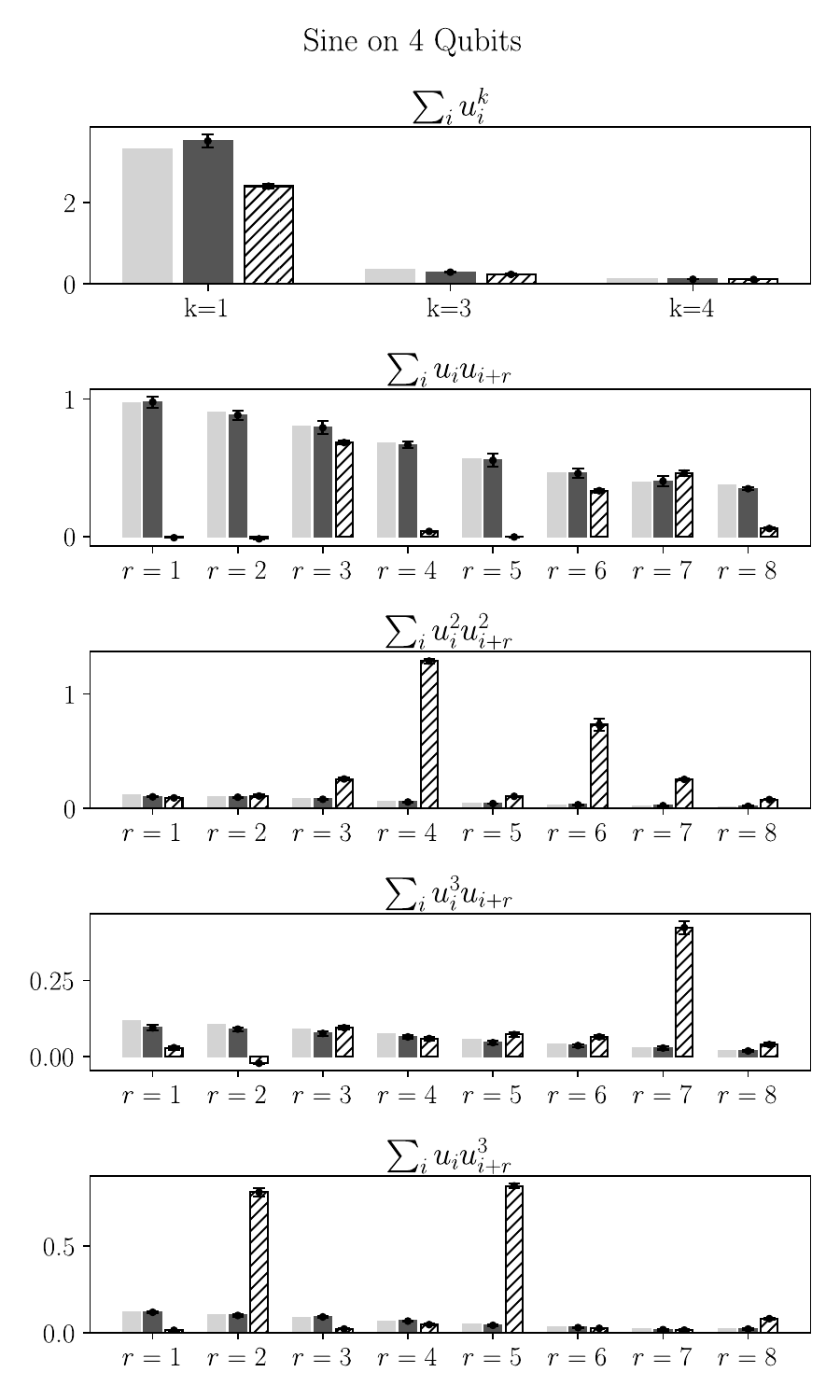}
        \label{fig:all_sums_sine4}
    \end{subfigure}
    \\
    \begin{subfigure}[t]{.445\textwidth}
        \centering 
        \includegraphics[width=\linewidth]{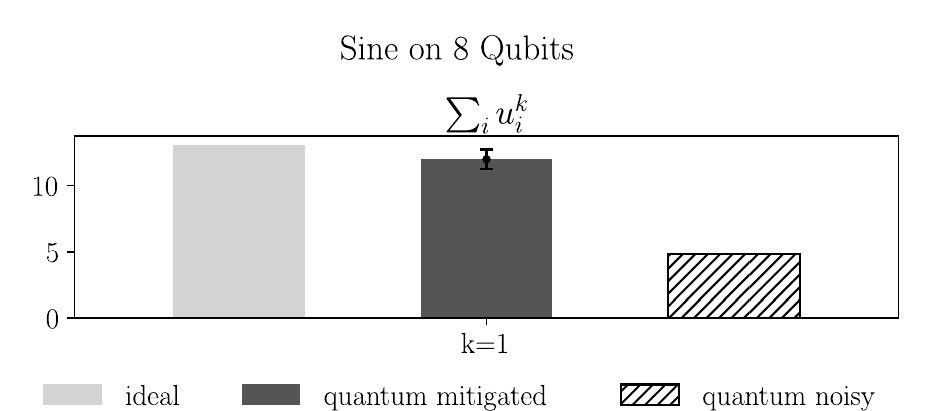}
        \label{fig:all_sums_sine8}
    \end{subfigure}
    \caption{Comparison of ideal values, error mitigated quantum measurements and noisy quantum measurements for the sine signal test case on 4 and 8 qubits.}
\end{figure}

\begin{figure}[ht!]
     \centering
     \begin{subfigure}[b]{.45\textwidth}
        \centering
        \includegraphics[width=\linewidth]{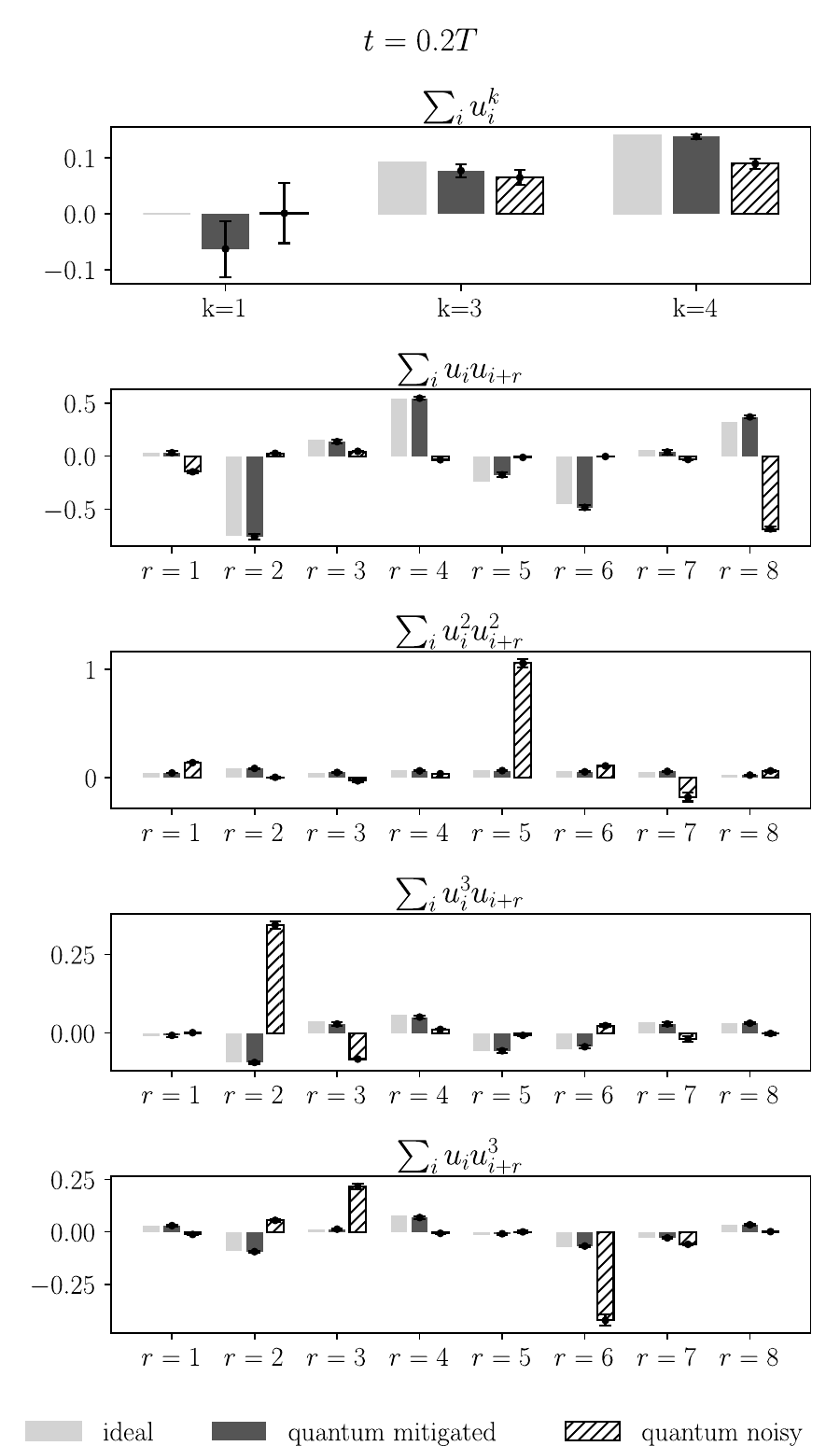}
        \caption{}
        \label{fig:all_sums_t1}
     \end{subfigure}
     \hfill
     \begin{subfigure}[b]{.45\textwidth}
         \centering
        \includegraphics[width=\linewidth]{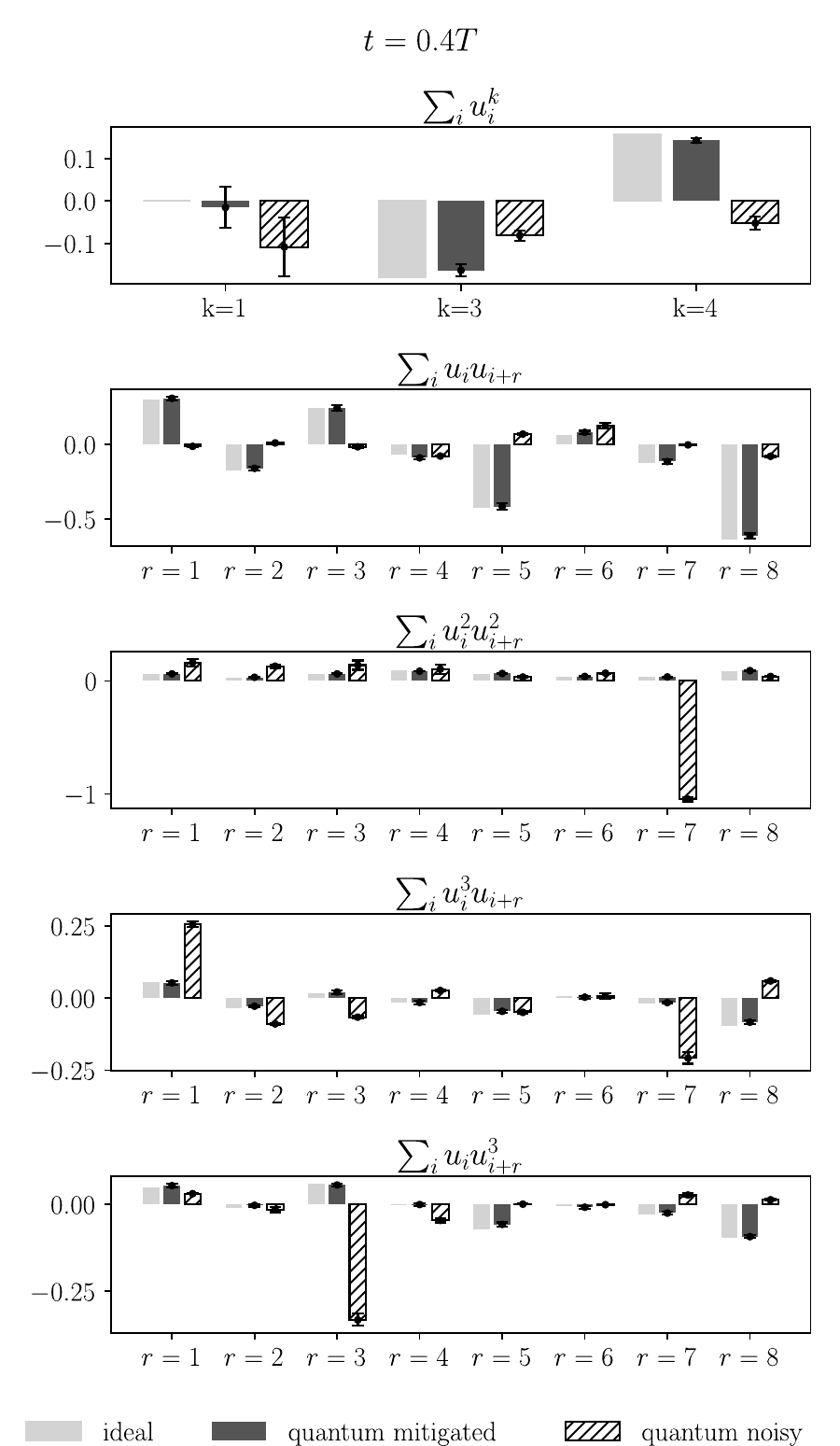}
        \caption{}
        \label{fig:all_sums_t2}
     \end{subfigure}
    \caption{Comparison of ideal values, error mitigated quantum measurements and noisy quantum measurements for the time steps $t=0.2T$ and $t=0.4T$ from the Burgers' evolution in Figure~\ref{fig:moments_over_time}.}
    \label{fig:all_sums_t1_t2}
\end{figure}

\clearpage

\begin{figure}[ht!]
     \centering
     \begin{subfigure}[b]{.45\textwidth}
        \centering
        \includegraphics[width=\linewidth]{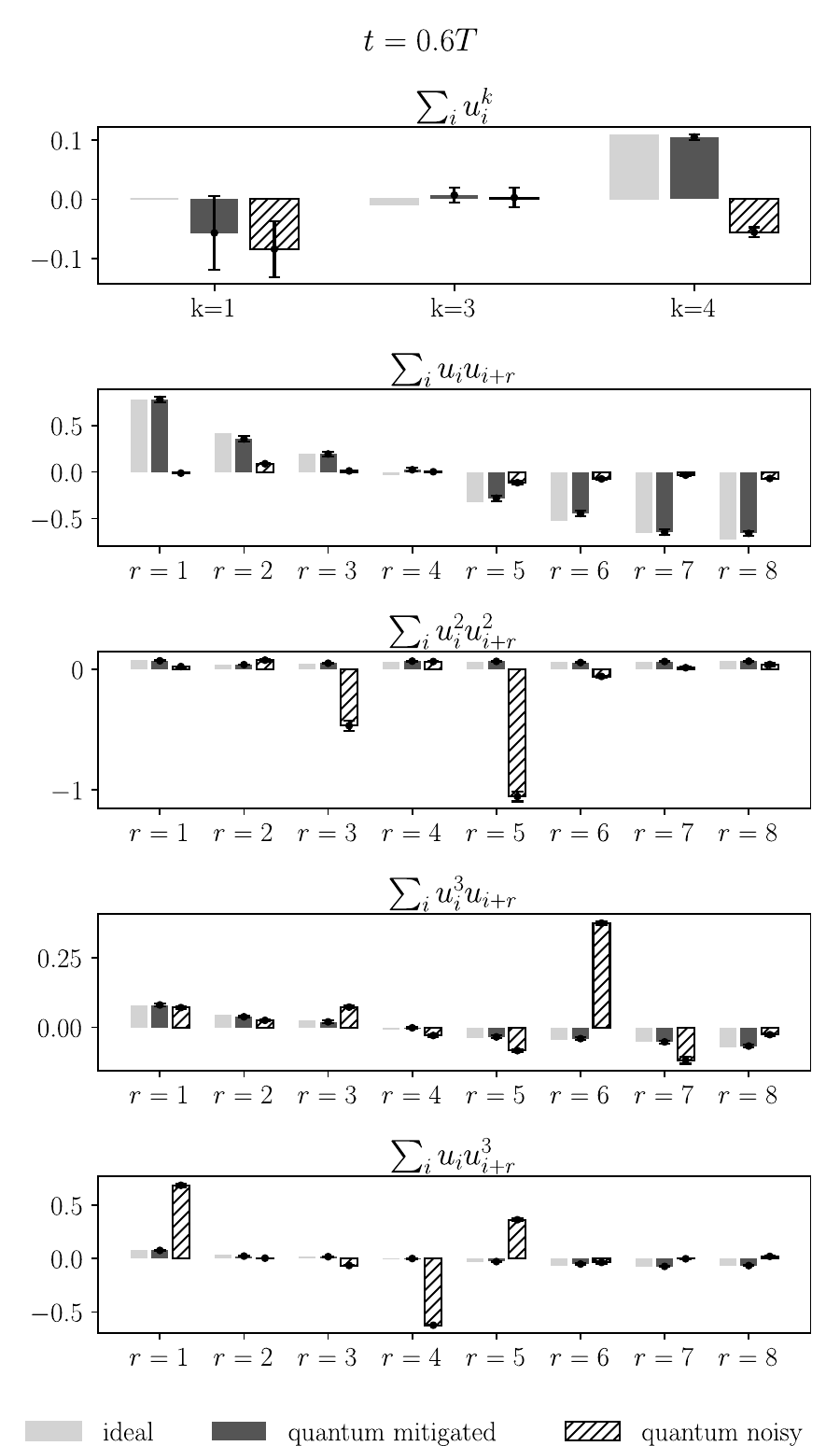}
        \caption{}
        \label{fig:all_sums_t3}
     \end{subfigure}
     \hfill
     \begin{subfigure}[b]{.45\textwidth}
         \centering
        \includegraphics[width=\linewidth]{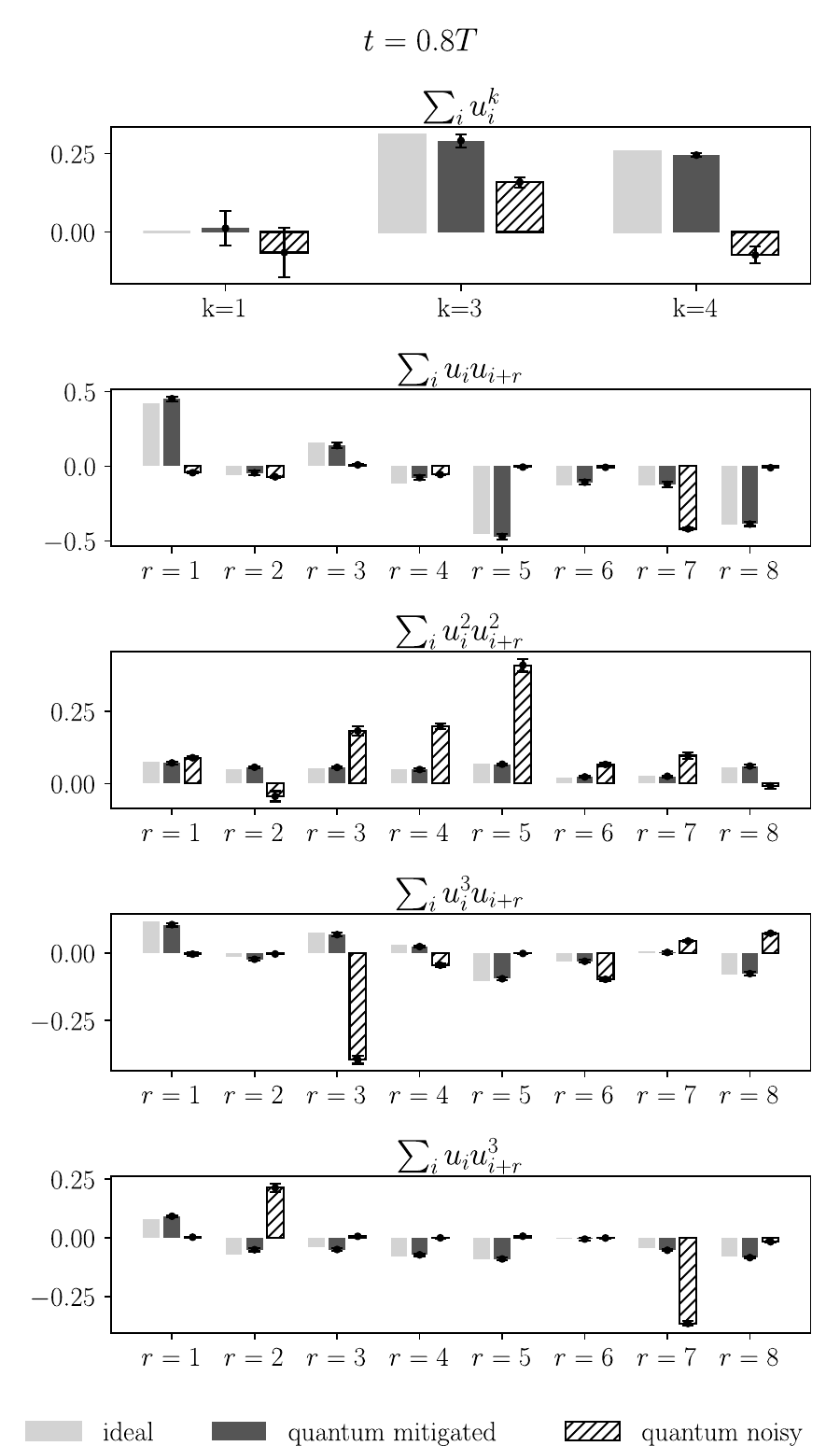}
        \caption{}
        \label{fig:all_sums_t4}
     \end{subfigure}
    \caption{Comparison of ideal values, error mitigated quantum measurements and noisy quantum measurements for the time steps $t=0.6T$ and $t=0.8T$ from the Burgers' evolution in Figure~\ref{fig:moments_over_time}.}
    \label{fig:all_sums_t3_t4}
\end{figure}

\section{QESEM (Quantum Error Suppression and Error Mitigation)}\label{sec:QESEM}

\noindent
The experiments presented in this work were performed using the QESEM software suite on IBM QPUs. Quantum hardware is intrinsically limited by gate infidelity (arising from, e.g., decoherence and calibration errors), which fundamentally constrains the depth and scale of executable circuits. The probability of running a circuit of \(V\) gates without a single error decays as \(\exp(-V \cdot IF)\), where \(IF\) denotes gate infidelity. In regimes where \(V \cdot IF \gtrsim 1\), expectation value estimates become dominated by systematic bias.

Error mitigation techniques that remove this bias do so at the cost of increased sampling complexity and QPU runtime. Unbiased estimators, such as those based on quasi-probabilistic decompositions, typically exhibit variance overhead scaling as \(\exp(\lambda\, IF\, V_a)\), where \(V_a\) is the ``active'' circuit volume (i.e., the gates within the causal light cone of the observable) and \(\lambda \in [2,4]\) encapsulates the protocol’s resource exponent. This scaling reflects a fundamental trade-off: while bias can be suppressed to arbitrary precision, the required sampling cost grows exponentially.

QESEM (Quantum Error Suppression and Error Mitigation) addresses this trade-off with provable convergence guarantees. Its end-to-end pipeline includes rapid, device-wide characterization to identify coherent errors and map native-gate fidelities; noise-aware transpilation that optimizes qubit mappings and gate decompositions to minimize active volume; coherent-error suppression via tailored gate recalibration and Pauli twirling; predictive re-characterization to fit a local Pauli error model for both gates and SPAM operations; and adaptive quasi-probabilistic error mitigation, which interleaves characterization and mitigation circuits to correct residual errors while tracking hardware drift.

QESEM provides a streamlined API that enables users to specify quantum circuits, observables, and target error tolerances in a hardware-agnostic fashion. The software handles device characterization, circuit transpilation, and adaptive error mitigation automatically, allowing practitioners to focus on algorithmic development rather than device-specific details. A typical workflow requires only a few lines of code, as illustrated below:

\definecolor{vsCommentGray}{gray}{0.6}       
\definecolor{vsKeyword}{RGB}{86,156,214}     

\lstset{
  language=Python,
  basicstyle=\ttfamily\small,                 
  frame=single,                               
  framerule=0.5pt,                            
  framesep=3pt,                               
  commentstyle=\color{vsCommentGray},         
  breaklines=true,                            
  keywordstyle=\bfseries\color{vsKeyword},    
  showstringspaces=false,                     
}

\begin{lstlisting}
import qiskit, qedma_api
from qiskit.quantum_info import SparsePauliOp

# Load circuit and observable
circ = qiskit.qasm2.load("example.qasm")
observable = SparsePauliOp("IIXIXI") 
qpu_time_limit_hrs = 0.5   

# Create and run job
job = qedma_client.create_job(
    circuit=circ,
    observables=[observable],
    precision=0.05,
    backend="ibm_fez",
    circuit_options=qedma_api.CircuitOptions(transpilation_level=1),
)
qedma_client.start_job(
    job_id=job.job_id
    max_qpu_time=datetime.timedelta(hours=qpu_time_limit_hrs)
)

# Get the results
job = qedma_client.get_job(job_id='123456789', include_results=True)
print(job.results)

\end{lstlisting}
For further details on QESEM, see~\cite{QESEM}).

\section{Dimensional Expressivity Analysis}\label{sec:DEA}
A dimensional expressivity analysis (DEA)~\cite{Funcke_2021} can reveal the number of redundant parameters in an ansatz and provide information about its local expressivity, that is, how well it parametrizes the accessible subspace of the Hilbert space around a given initialization point. In particular, the Jacobian is defined as the real derivative of the state in Equation~(\ref{eq:parameterized_ansatz}) with respect to the ansatz parameters $\theta$, and its rank corresponds to the number of independent directions in which the circuit can vary the output state. For a real-valued quantum circuit acting on $n$ qubits, this rank is bounded above by $2^n - 1$, corresponding to the dimension of the real projective Hilbert space (excluding normalization).

\begin{figure}[!htbp]
  \centering
  \includegraphics[width=\textwidth]{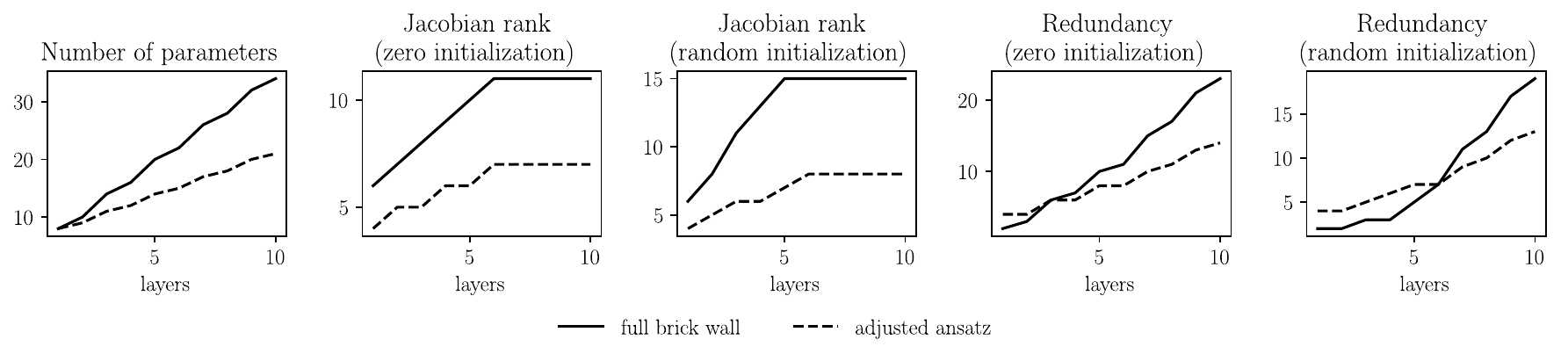}
  \caption{Dimensional expressivity analysis of the full brick-wall-structured ansatz and the hardware-adjusted ansatz.}
  \label{fig:naive_vs_adjusted_ansatz}
\end{figure}

In our case, the maximum attainable rank is 15, as the circuit only produces real amplitudes on 4 qubits. As can be seen in Figure~\ref{fig:naive_vs_adjusted_ansatz}, the full brick wall ansatz reaches this upper bound when the number of layers is sufficiently large, suggesting that it fully spans the accessible real subspace. In contrast, the adjusted ansatz saturates at a Jacobian rank of 7, indicating parameter redundancy. Nevertheless, this level of expressivity is sufficient to capture a wide range of relevant states, while using fewer parameters than the full brick wall ansatz. 

One identifiable source of redundancy in the adjusted ansatz is the presence of two consecutive $R_Y$ gates on qubits 2 and 4 at the end of the circuit. Since $R_Y(\theta_1)R_Y(\theta_2)=R_Y(\theta_1+\theta_2)$, a single parameter can realize the same transformation, allowing the ansatz to achieve the desired transformations efficiently. Additionally, the overall entangling structure also influences the circuit's expressivity. Alternating CNOT directions or introducing SWAP operations can increase the Jacobian rank and thereby enhance the expressivity of the circuit. 

While such redundancies may appear inefficient, overparameterization can in fact enhance the trainability of variational quantum algorithms. Redundant parameters may help smooth the optimization landscape, mitigate the impact of hardware noise, and increase the chances of locating suitable minima~\cite{Kim_2021, Fontana_2021, Funcke_2021}. 

To further evaluate the practical expressivity of the adjusted quantum circuit on 4 qubits, an additional test was performed in which the circuit parameters were optimized to reproduce randomly sampled target states from the corresponding Hilbert space. The optimization minimized the cosine distance, which measures the deviation in orientation between two vectors while being invariant under global amplitude scaling. The cost function $C$ was defined as 
\begin{equation}
    C = 1 - \frac{\braket{\phi_{\text{target}}, \phi_{\text{prepared}}}}{\|\phi_{\text{target}}\| \cdot \|\phi_{\text{prepared}}\|}.    
\end{equation}
Although the DEA rank does not reach the maximum, the optimization demonstrates that the adjusted circuit can accurately represent the target states, achieving cosine distances on the order of $10^{-11}$. This shows that the ansatz is highly effective in practice, despite theoretical reduction in expressivity.

These results indicate that the adjusted ansatz already captures the relevant state space efficiently. Still, targeted architectural modifications such as adding entangling layers or redistributing parameters could be explored in future work to enhance its expressivity even further.

\end{appendices}

\end{document}